\documentclass[aps,prl,showpacs,superscriptaddress,twocolumn,nofootinbib]{revtex4-1}

\usepackage{lmodern}

\usepackage[latin9]{inputenc}
\usepackage{amsmath}
\usepackage{amssymb}
\usepackage{graphicx}
\usepackage{epstopdf}
\usepackage{color}
\usepackage{enumerate}

\begin{document}

\title{Low Magnetic Field Regime of a Gate-Defined Constriction in High-Mobility Graphene}

\author{L. Veyrat}
\author{A. Jordan}
\author{K. Zimmermann}
\author{F. Gay}
\affiliation{Univ. Grenoble Alpes, CNRS, Grenoble INP, Institut N\'{e}el, 38000 Grenoble, France}
\author{K. Watanabe}
\affiliation{National Institute for Materials Science, 1-1 Namiki, Tsukuba 306-0044, Japan}
\author{T. Taniguchi}
\affiliation{National Institute for Materials Science, 1-1 Namiki, Tsukuba 306-0044, Japan}
\author{H. Sellier}
\author{B. Sac\'{e}p\'{e}}
\affiliation{Univ. Grenoble Alpes, CNRS, Grenoble INP, Institut N\'{e}el, 38000 Grenoble, France}
\email{benjamin.sacepe@neel.cnrs.fr}

\begin{abstract}

\bf{We report on the evolution of the coherent electronic transport through a gate-defined constriction in a high-mobility graphene device from ballistic transport to quantum Hall regime upon increasing the magnetic field. At low field, the conductance exhibits Fabry-P\'{e}rot resonances resulting from the npn cavities formed beneath the top-gated regions.
Above a critical field $B^*$ corresponding to the cyclotron radius equal to the npn cavity length, Fabry-P\'{e}rot resonances vanish and snake trajectories are guided through the constriction with a characteristic set of conductance oscillations. Increasing further the magnetic field allows us to probe the Landau level spectrum in the constriction, with distortions due to the combination of confinement and de-confinement of Landau levels in a saddle potential. These observations are confirmed by numerical calculations.
}

\end{abstract}
\maketitle

Controlling electron transport with gate-defined constrictions is key in many quantum coherent experiments. In two-dimensional electron gases formed in semiconducting heterostructures, constrictions are commonly made with nano-patterned split-gate electrodes that draw a short and narrow channel of conduction acting as a quantum point contact (QPC)~\cite{Beenakker11}. Such a local electrostatic gating enables fine tuning of the electron transmission through the QPC with the ensuing quantization of the conductance, both in the ballistic~\cite{Wharam88,Wees88} and in the quantum Hall regimes under strong magnetic field~\cite{Wees89,Kouwenhoven90}.

In graphene, engineering such a gate-defined QPC has proven difficult due to the absence of band gap for the Dirac electrons~\cite{Castro2009}, which prevents the formation of insulating regions below gate electrodes ~\cite{Huard2007,Nakaharai2011,Xiang16,Zimmermann2017} (depleting an electron (hole)-doped graphene indeed leads to a hole (electron)-doped region beneath gate electrodes). At zero magnetic field, the resulting npn junction is partially transparent to charge carriers, thereby hampering the realization of gate-defined QPCs in graphene. Still, other approaches based on etched constrictions~\cite{Tombros11,Terres16} or gate-defined long one-dimensional channels~\cite{Kim2016} have evidenced conductance quantization at zero magnetic field. Besides, the bilayer graphene case offers more flexibility as a gap can be induced with an out-of-plane electric field enabling field effect depletion and conductance quantization in a split-gate geometry~\cite{Li2018,Overweg18a,Eich18a,Eich18b,Overweg18b,Overweg18c}.

Recently, some of us demonstrated that, in the quantum Hall regime of high-mobility graphene, the gap that opens at the charge neutrality point between electron and hole-type broken symmetry states enables to operate a gate-defined QPC with fine-tuning of the edge-channels transmission~\cite{Zimmermann2017}.
This result, obtained with hBN-encapsulated graphene in van der Waals heterostrutures, opens the way to more elaborated quantum devices based on local top or bottom gating, towards exploring quantum coherent phenomena in graphene and new correlated fractional quantum Hall states~\cite{Du09,Bolotin09,Dean11,Amet14,Zibrov2017} with quantum Hall interferometry~\cite{Wees89b,Ji03}, non-equilibrium measurements~\cite{Altimiras10}, shot noise experiments~\cite{Saminadayar97, Heiblum97} or electron quantum optics~\cite{Bocquillon14}. 

However, the impact of the split-gate geometry on the ballistic transport at zero magnetic field, and the transition to the high-field quantum Hall regime where transport is dominated by the constriction, remain to be investigated. Understanding the behavior over the whole magnetic field range is necessary for designing future of split-gated graphene devices. 

In this work, we show how the transport through a split-gated constriction in high mobility graphene evolves from a low-field regime dominated by Fabry-P\'erot (FP) interference~\cite{Young2009,Gu2011,Nam2011,Campos2012,Grushina2013,Rickhaus2013,BenShalom16,Oksanen2014,Rickhaus2015} under the top gates, into a high-field regime dominated by the constriction.
At low field, the analysis of the FP interference together with self-consistent electrostatic simulations enable to assess the cavity length and estimate the potential profile across the n-p-n junction and inside the constriction. 
 At a critical field $B^*$, we identify a transition from conduction below the top-gates to transport toward the constriction.
Above $B^*$, the transport is mediated by snake trajectories which guide electrons toward the constriction. This regime is characterized by the emergence of snake oscillations, validated by numerical calculations, and of Landau levels of the constricted region. The confinement-de-confinement effect of the saddle potential modifies the Dirac Landau level spectrum in the constriction, leading to non-linear field and gate dispersions that we model theoretically.

\begin{figure}
	\centering
	\includegraphics[width=1\columnwidth]{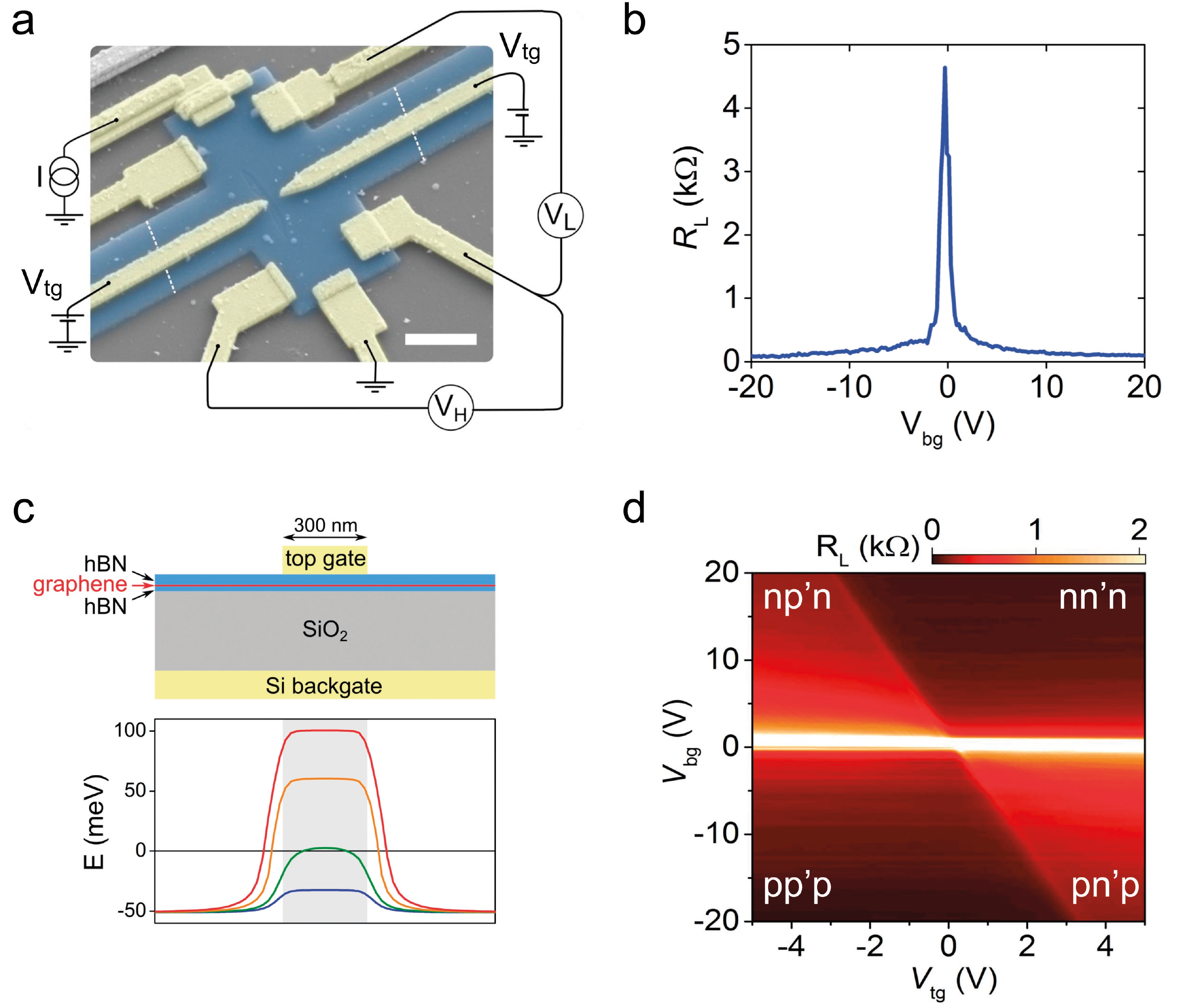}
	\caption[width=\columnwidth]{(a) Scanning electron micrograph of the split-gated hBN-graphene-hBN device showing the measurement configuration. $V_L$ and $V_H$ correspond to the longitudinal and Hall voltages, respectively. White dotted lines indicate the edge of the graphene flake under the top hBN flake. The scale bar is 1~$\mu$m. (b) Longitudinal resistance $R_{\text{L}}$ versus back-gate voltage $V_{\text{bg}}$ at zero top-gate voltage. (c) Schematic of the heterostructure and simulated energy profiles for different top-gate voltages (see Supplementary Information section 1). (d) Colormap of $R_{\text{L}}$ versus $V_{\text{bg}}$ and $V_{\text{tg}}$, showing the four regions of different charge configurations.}
	\label{fig1}
\end{figure}

High mobility graphene heterostructures are made of an exfoliated graphene monolayer encapsulated between two hexagonal boron nitride (hBN) flakes via the van-der-Waals pick-up technique~\cite{Wang2013}. In this study, we focus on a graphene device encapsulated between $18\,$nm/$56\,$nm thick bottom/top hBN flakes sitting on a SiO$_2$/Si++ substrate that serves as a back-gate electrode. Suitable etching allows for one-dimensional edge contacts to be deposited on the edges of the hBN/graphene/hBN stack~\cite{Wang2013}. Top-gate electrodes are patterned onto the top hBN flake in a split-gate geometry~\cite{Zimmermann2017}, with a separation of $200\,$nm and a gate width of $300\,$nm measured by SEM micrograph. Six ohmic contacts enable measurements of Hall and longitudinal resistances $R_{\rm H}$ and $R_{\rm L}$ in 4-terminal configuration, as shown schematically in Fig.~\ref{fig1}a. All measurements are performed at a temperature of 4.2~K.

We begin this study by discussing the device response to gate voltages and the signature of ballistic transport at zero magnetic field. Figure~\ref{fig1}b shows the back-gate voltage dependence of the longitudinal resistance $R_{\text{L}}$ that exhibits a sharp resistance peak corresponding to the Dirac point of the graphene flake. From the Drude conductance, we extract a Hall mobility of 85000 cm$^2$V$^{-1}$s$^{-1}$, which corresponds to a mean free path of 1~$\mu$m at a charge carrier density $n \approx 1\times 10^{12}$ cm$^{-2}$. This value of the mean free path indicates that transport is ballistic across a distance far superior to the width of the top-gated region.
The longitudinal resistance $R_{\text{L}}$ versus back-gate voltage $V_{\text{bg}}$ and top-gate voltage $V_{\text{tg}}$ is shown in Fig.~\ref{fig1}d. The charge neutrality point of the bulk graphene is apparent as the horizontal line independent of $V_{\text{tg}}$ at $V_{\text{bg}}^{\text{CNP}} = 0.5$ V, indicating very little intrinsic doping. The diagonal ridge corresponds to a second peak in resistance, signaling the charge neutrality point of the graphene top-gated region, which is under the electrostatic influence of both the back-gate and top-gate electrodes. From its slope, we extract a top-gate capacitance $C_{\text{tg}}=65$~nF/cm$^2$, knowing the back-gate capacitance $C_{\text{bg}}=10$~nF/cm$^2$ from Hall measurements. These two resistance ridges define four regions of different polarities (nn'n, np'n, pp'p, pn'p). This map at zero magnetic field shows no indication of the presence of the constriction (such as conductance quantization) due to the small size of the constriction as compared to the much wider regions below the top gates which are always conducting even in the np'n and pn'p bipolar regimes~\cite{Nakaharai2011,Xiang16, Zimmermann2017}.

\begin{figure}[t]
	\centering
	\includegraphics[width=1\columnwidth]{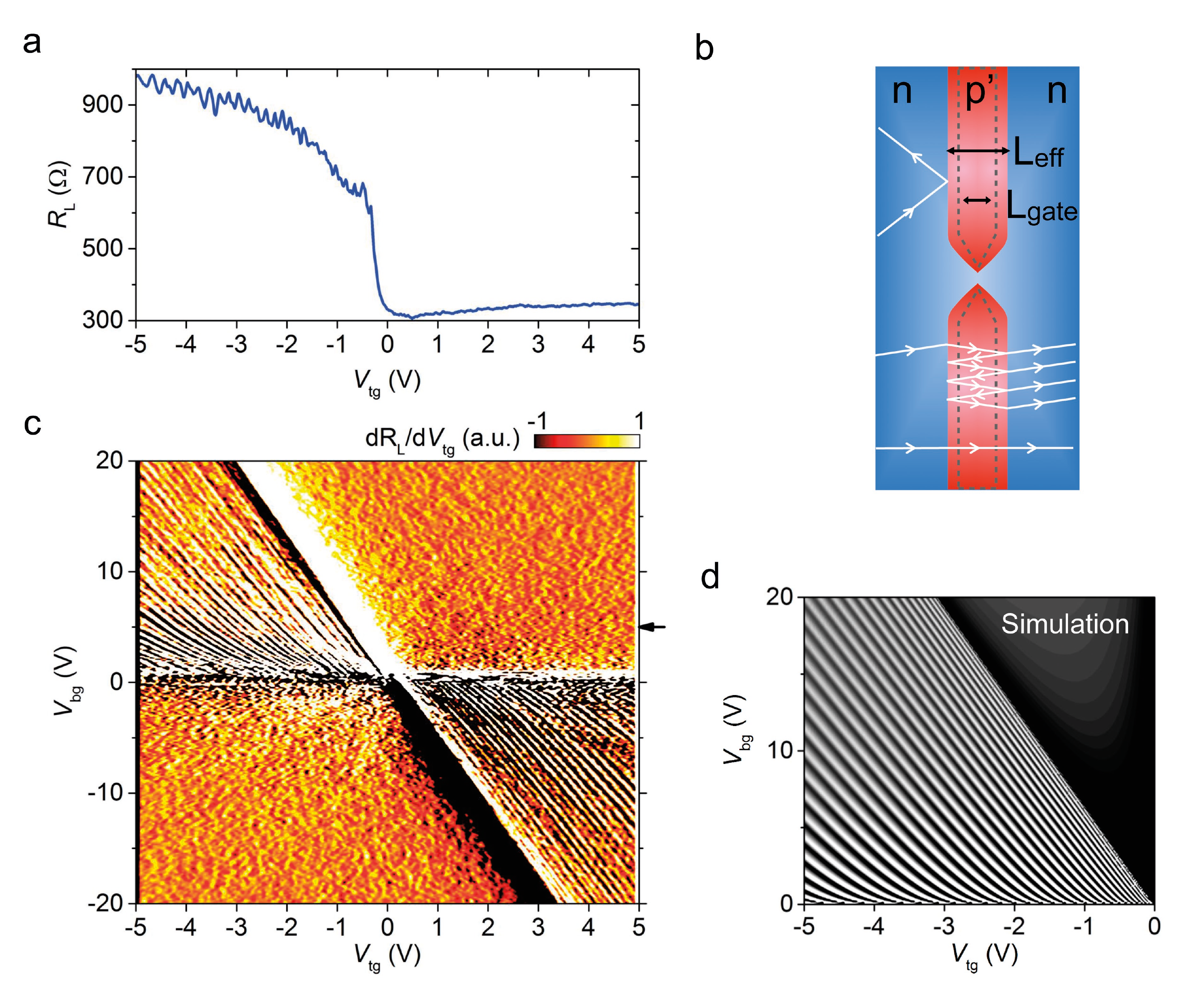}
	\caption{(a) Longitudinal resistance $R_{\text{L}}$ versus top-gate voltage $V_{\text{tg}}$ at $V_{\text{bg}}=5$~V. (b) Schematic of ballistic electron trajectories through the top-gated region. While normally incident electrons are perfectly transmitted (Klein tunneling) and large angle electrons are reflected, electrons with intermediate angles produce Fabry-P\'{e}rot interference. $L_{\text{eff}}$ and $L_{\text{gate}}$ represent the effective Fabry-P\'{e}rot cavity length and physical top-gate width, respectively. (c) Derivative of the longitudinal resistance d$R_{\text{L}}$/d$V_{\text{tg}}$ as a function of $V_{\text{bg}}$ and $V_{\text{tg}}$. The black arrow indicates the position of the curve in (a). (d) Derivative of the simulated longitudinal resistance (see Supplementary Information section 2).}
	\label{fig2}
\end{figure}

Inspecting the $V_{\text{tg}}$-dependence of the resistance shown in Fig.~\ref{fig2}a for a fixed back-gate voltage, we see that prominent pseudo-periodic oscillations of the longitudinal resistance emerge in the bipolar regime with a pseudo-period $\Delta V_{\text{tg}}$ that increases on more negative $V_{\text{tg}}$ values and a visibility of $\sim 3\%$ over the range $V_{\text{tg}} = $[-$5\text{V}$, -$2\text{V}]$. The derivative d$R_{\text{L}}$/d$V_{\text{tg}}$ over the entire resistance map in Fig.~\ref{fig2}c reveals their presence in the two bipolar regimes and shows their dispersion with $V_{\text{bg}}$ and $V_{\text{tg}}$.

Such resistance oscillations are similar to that observed in single or double pn-junction devices \cite{Young2009,Gu2011,Nam2011,Campos2012,Grushina2013,Rickhaus2013,BenShalom16,Oksanen2014,Rickhaus2015} and provide direct signature of quantum interference of ballistic electron trajectories bouncing between the two pn junctions of the top-gated region. As illustrated in Fig.~\ref{fig2}b, the two electrodes of the split-gate form a constriction but also FP cavities beneath their long section with parallel edges. There, the two parallel and partially transmitting pn junctions induce multiple interference for ballistic electrons and thus FP resonances in the transmission of the resulting np'n (pn'p) cavity~\cite{Katsnelson2006,Cheianov2006PRB,Shytov2007,Masir2010}. After averaging over all incidence angles, these sharp resonances translate into smooth resistance oscillations such as those visible in Fig.~\ref{fig2}a.

\begin{figure*}[t!]
    \centering
    \includegraphics[width=0.9\linewidth]{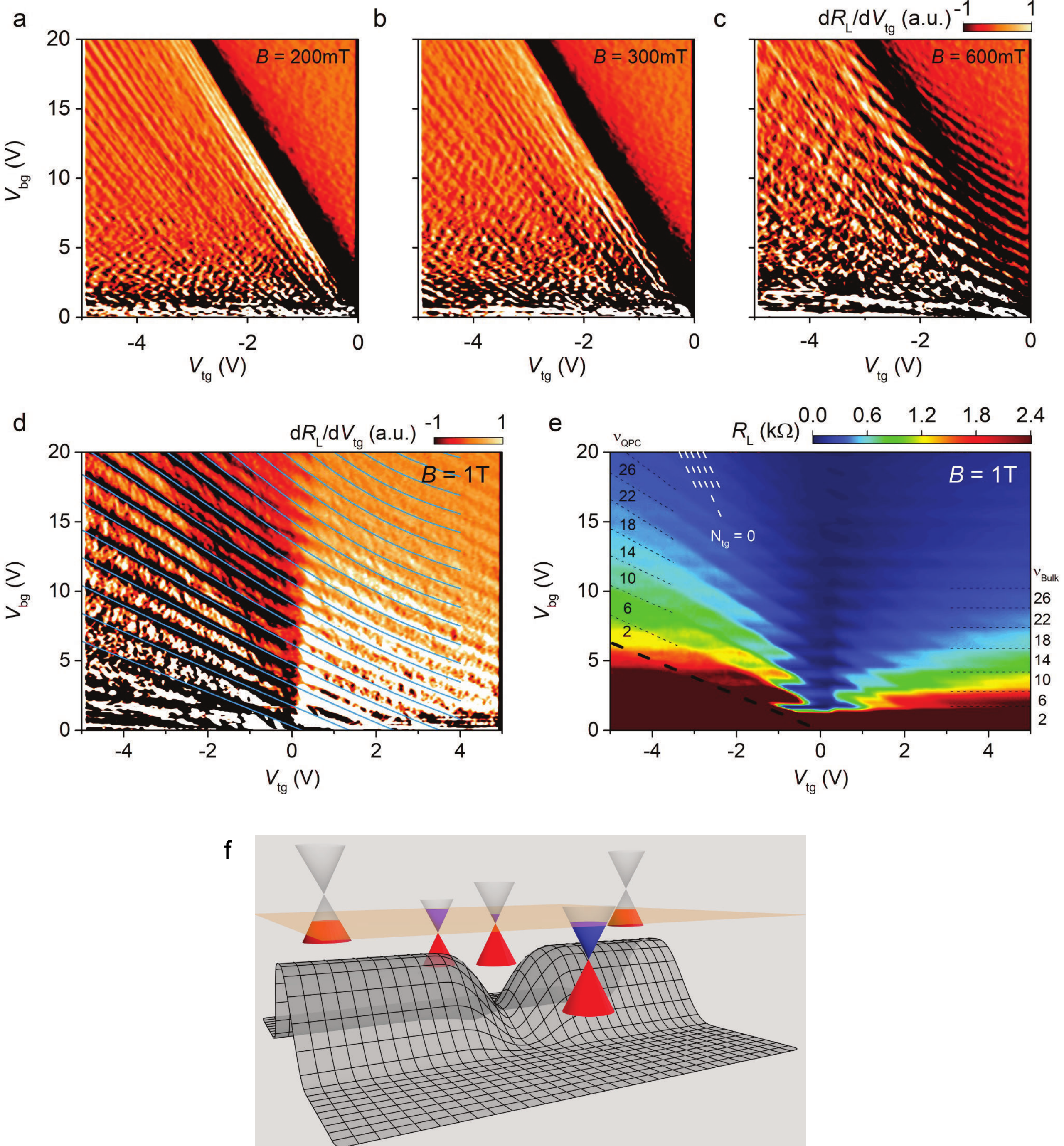}
    \caption{(a-d) Derivative of the longitudinal resistance d$R_{\text{L}}$/d$V_{\text{tg}}$ versus back-gate voltage $V_{\text{bg}}$ and top-gate voltage $V_{\text{tg}}$ at $B = 0.2$~T, 0.3~T, 0.6~T, and 1~T. In (d), the blue lines correspond to the fitted Landau levels. (e) Longitudinal resistance $R_{\text{L}}$ at $B=1$~T. The pattern is typical of edge-channel transport in the quantum Hall regime through a split-gated device. Dashed lines are guides for the eye showing the limits of regions with different filling factors $\nu_{\text{b}}$ and $\nu_{\text{QPC}}$ in the bulk and constricted regions, respectively. The thick dashed line corresponds to the charge neutrality of the constricted area. The Landau levels in the top-gated region are represented by the white dashed lines. (f) Schematic of the saddle potential at the constriction. The Dirac cones indicate the chemical potential in the top-gated, bulk, and constricted zones.}
    \label{fig3}
\end{figure*}

Interestingly, despite the fact that the resonant cavity is cut in two parts due to the split-gate geometry, FP oscillations are observed with a significant visibility in the entire area of the bipolar regime under the top gates (Fig.~\ref{fig2}c). At low back-gate voltage, far from the diagonal ridge, the constriction is also in the bipolar regime, but forms a narrow non-resonant cavity with non-parallel pn interfaces, and the resonances average out to zero.
At higher back-gate voltage, close to the diagonal ridge, the constriction is in the unipolar regime and cannot produce interference, but the measured resistance pattern is almost unperturbed.
This indicates that the constriction region has a virtually negligible contribution to the total conductance at zero magnetic field.
 
Quantitative analysis of the FP oscillations enables to extract an effective cavity length $L_{\text{eff}}$ between the two pn junctions, which depends on the applied top-gate and back-gate voltages. The pseudo-period in charge carrier density is given by $\Delta n_{\rm tg} = 2\sqrt{\pi n_{\rm tg}}\,/\,L_{\text{eff}}$, where $n_{\rm tg}$ is the charge carrier density beneath the top-gate~\cite{Young2009,Grushina2013}. At $V_{\text{bg}}=10$~V, one can calculate the cavity length by estimating the pseudo-period $\Delta n_{\rm tg}$ over several oscillations. By averaging the pseudo-period between $V_{\text{tg}}=-5$~V and $-1.5$~V, we find $L_{\text{eff}} \simeq 380$~nm, which is slightly larger than the width of the top-gate electrodes. Self-consistent electrostatic simulations (see Supplementary Information section 1) have been performed to calculate the potential profile across the npn junction (Fig.~\ref{fig1}c), and a cavity length of 366~nm has been obtained for $V_{\text{bg}}=10$~V and $V_{\text{tg}}=-3$~V, which is close to the value extracted from the data at the same gate voltages.

We furthermore complement this analysis by numerical simulations of the transmission through the npn junctions. We calculate the conductance by averaging the angular dependence of the transmission obtained within the WKB approximation~\cite{Shytov2007} at zero magnetic field, and using a potential profile determined by self-consistent simulations (see Supplementary Information section 2).
The resulting gate-voltage dependence of the transmission shown in Fig.~\ref{fig2}d reproduces qualitatively the experimental pattern of resistance oscillations. The curvature of the interference fringes at low back-gate voltage corresponds to a significant increase of the cavity length, due to a weaker screening of the top-gate voltage by the low carrier density in the bulk region.

 \begin{figure*}[t]
    \centering
    \includegraphics[width=0.9\linewidth]{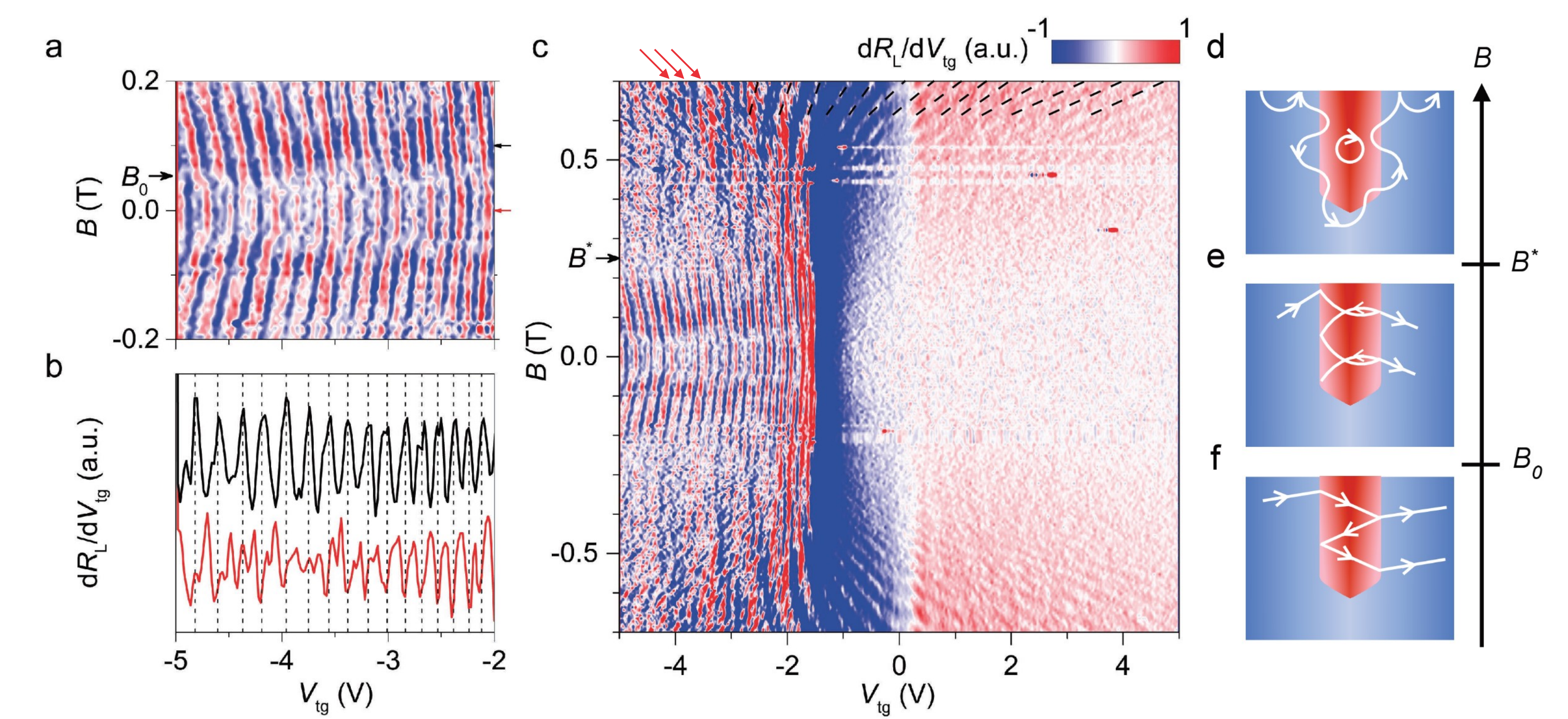}
    \caption{(a) Derivative of the longitudinal resistance d$R_{\text{L}}$/d$V_{\text{tg}}$ as a function of the magnetic field $B$ and top-gate voltage  $V_{\text{tg}}$, at $V_{\text{bg}} = 10\, \text{V}$. Fabry-P\'{e}rot oscillations are bent in the presence of magnetic field, and undergo a phase shift at $B=B_0$. (b) Line-cuts at 0 and 0.1~T, as indicated by arrows on the right axis in (a). (c) Same data as in (a) over a larger range of $V_{\text{tg}}$ and $B$. The Fabry-P\'{e}rot oscillations are only visible at low field in the bipolar regime. Dashed lines on the top are guide for the eye for the Shubnikov-de-Haas oscillations from the constricted area. The red arrows point at small amplitude snake oscillations. (d-f) Schematic of the ballistic electron trajectories within the upper top-gated region upon increasing magnetic field. (f) At zero field, trajectories are straight. (e) Above $B_0$, trajectories form loops, picking-up the graphene Berry phase. (d) Above $B^*$, electrons form closed orbits and no longer reach the opposite side of the junction. Snake orbits appear along the pn interface, guiding the electronic trajectories through the constriction.}
    \label{fig4}
\end{figure*}

We now show that on increasing the magnetic field towards the quantum Hall regime, the contribution of the constricted region emerges and becomes predominant in the conductance. Given the electronic mobility $\mu = 85000$~cm$^2$V$^{-1}$s$^{-1}$, Landau levels should appear roughly above a field such that $B=1/\mu=0.12$~T. Figure~\ref{fig3} presents the dependence of the resistance oscillations with back-gate and top-gate voltages, at different magnetic fields. At $B=0.2$~T (Fig.~\ref{fig3}a), the FP oscillations are visible in the bipolar regime. From $B=0.6$~T on, another set of parallel lines appears, extending from the unipolar to the bipolar regime, with a smaller slope than that of the Dirac ridge of the top-gated area (Fig.~\ref{fig3}c,d). Since Landau levels in the bulk graphene should manifest as horizontal lines independent of the top-gate voltage, and those in the top-gated area should be parallel to the diagonal Dirac ridge, these new lines hence correspond to a region with an intermediate capacitive coupling, which is the constriction region located between the two electrodes of the split-gate as previously identified in Ref.~\cite{Zimmermann2017}.

At $B=1$~T in Fig.~\ref{fig3}d,  we extract from the slope of the line passing through zero top-gate and back-gate voltages a capacitance ratio $C^{\text{QPC}}_{\text{tg}}/C_{\text{bg}}=1.2$ ($C^{\text{QPC}}_{\text{tg}}$ is the capacitance between the graphene constriction region and the top-gate), a value close to the value obtained from a self-consistent electrostatic simulation of the split-gate geometry (see Supplementary Information section 1).
Although  Landau level are usually linear with gate voltage, non-linearities can be observed in the gate dependence of the constriction's Landau levels. This is a consequence of the saddle potential in the constriction and will be further discussed below.
Knowing the capacitive coupling of the constriction, one can calculate the number of Landau levels expected at $B=1$~T for $V_{\text{bg}}=10$~V and $V_{\text{tg}}$ varying from $-5$ to 5~V, corresponding to a Fermi energy change in the constriction from 62 to 118 meV. Given the graphene Landau quantization $\epsilon_N=v_F\sqrt{2 \hbar e B N}$, $N$ being the Landau level index, this corresponds to a change from 3 to 11 filled Landau levels,  which is very close to the 10 oscillations observed over this range in Fig.~\ref{fig3}d. This analysis confirms that the new set of oscillations in Fig.~\ref{fig3}c-d are the Shubnikov-de-Haas oscillations from the constricted region.

Notice that at $B=0.2$~T, no Shubnikov-de-Haas oscillations from the top-gated area are visible, because their spacing in top-gate voltage $\delta V_{\text{tg}}$ is smaller than the gate resolution of the measurement. Furthermore, at $B=0.6$~T, the electron trajectories do not cross the top-gated area due to the formation of snake states as discussed in details in the following, so that the oscillations of the density of states in this region do not impact the transmission.
 
These emerging Shubnikov-de-Haas oscillations of the constricted region exhibit unusual features. At zero top-gate voltage in Fig.\ref{fig3}d, the Landau levels are equally spaced in density, as expected for a graphene sheet with a uniform electrostatic potential. At finite top-gate voltage however, the lines marking the Landau levels are not parallel to each others, indicating a gate-voltage-dependent level spacing. The spacing increases significantly for negative voltages and decreases slightly for positive voltages. A given Landau level therefore appears in the map as a curved line, with an increasing (negative) slope for increasing Landau level index and a small upward curvature.

The origin of these deviations from equally spaced Landau levels result from the non-uniform electrostatic potential in the constriction. The split-gate electrode indeed creates a saddle potential with a strong curvature in the transverse direction and a weak curvature in the longitudinal direction (see Fig.\ref{fig3}f). For positive back-gate voltage, a negative split-gate voltage induces a transverse confinement that increases the magnetic confinement and leads to an increased Landau level spacing. Conversely, a positive split-gate voltage induces a transverse deconfinement that reduces the magnetic confinement and leads to a reduced Landau level spacing. A similar de-confinement effect has already been observed in single-top-gate graphene devices (without constriction). It results in a collapse of the Landau levels into a continuum of states for sufficiently large potential curvatures \cite{Gu2011}. Here, the collapse does not occur due to the presence of the longitudinal curvature, which is of opposite sign to the transverse one and thus restores the existence of closed orbits.

The theoretical equations for the Landau levels in a saddle potential have no analytical solution in the case of graphene where the dispersion relation is linear (the high magnetic field limit is not applicable here) \cite{Floser2010}. However, we can calculate an approximative expression for the Landau levels in the constriction, taking advantage of the formal equivalence between the linear and quadratic dispersion relations in case of an identical carrier density distribution \cite{Gu2011}, and given that an exact solution exists for the quadratic case \cite{Fertig87,Buttiker90} (see Supplementary Information section 6). The position of the Landau levels given by this approximate expression is shown as blue lines in Fig. 3d (in the range of validity of the solution \cite{Fertig87}). The two fitting parameters are the transverse and longitudinal curvatures of the density profile, and the obtained values are consistent with our self-consistent electrostatic simulations (see Supplementary Information section 6). This good agreement confirms that the observed Landau levels originate from the constriction. The discrepancies observed at large positive split-gate voltage, where the calculated levels curve more than in the measurement, could be due to the second-order approximation of the density profile in the constriction, which is in reality bell-shaped on a larger scale : when the cyclotron radius exceeds a few hundred nanometers, as it does at high density or at low field, the solution for a infinitely-parabolic saddle potential is not valid anymore, and the confinement effects tend to decrease \cite{Gu2011}.

At $B=1$~T, the graphene is already in the quantum Hall regime resulting in plateaus in the  longitudinal resistance with a parallelogram shape delimited by lines of equal filling factors in the bulk graphene (horizontal lines) and in the constricted region (the new diagonal lines discussed above), as displayed in Fig.~\ref{fig3}e.
The horizontal strips of constant bulk filling factor $\nu_{\text{b}}$ and the diagonal strips of constant filling factor in the constriction $\nu_{\text{QPC}}$ are indexed in Fig.~\ref{fig3}e. 
These parallelogram shape plateaus thus indicate that electron transmission is controlled by the number of edge channels passing through the constriction.
In the quantum Hall regime, the transmission through the device is due to edge channels passing through the constriction. Consequently, there is no feature parallel to the charge neutrality point of the top-gated area, that would indicate variations of the top-gated area filling factor $\nu_{\text{tg}}$.
At the magnetic field considered here, which is too small for broken symmetry state to develop and to open a gap between electron and holes, the exact value of the resistance within each plateau results from current equilibration between co-propagating edge channels along the pn interfaces~\cite{Nakaharai2011,Xiang16,Zimmermann2017}. 
In this case, as shown in Ref.~\cite{Zimmermann2017}, the conduction through the system is governed by three filling factors : $\nu_{\rm b}$, $\nu_{\rm QPC}$, and also $\nu_{\rm g}$ below the top gates, with different equilibration rules in the unipolar and bipolar regimes of the top gates. Notice that $\nu_{\rm QPC}=0$ occurs below the thick dashed line in Fig.\ref{fig3}e, which represents the charge neutrality point of the constriction.

\begin{figure*}[t!]
    \centering
    \includegraphics[width=1\linewidth]{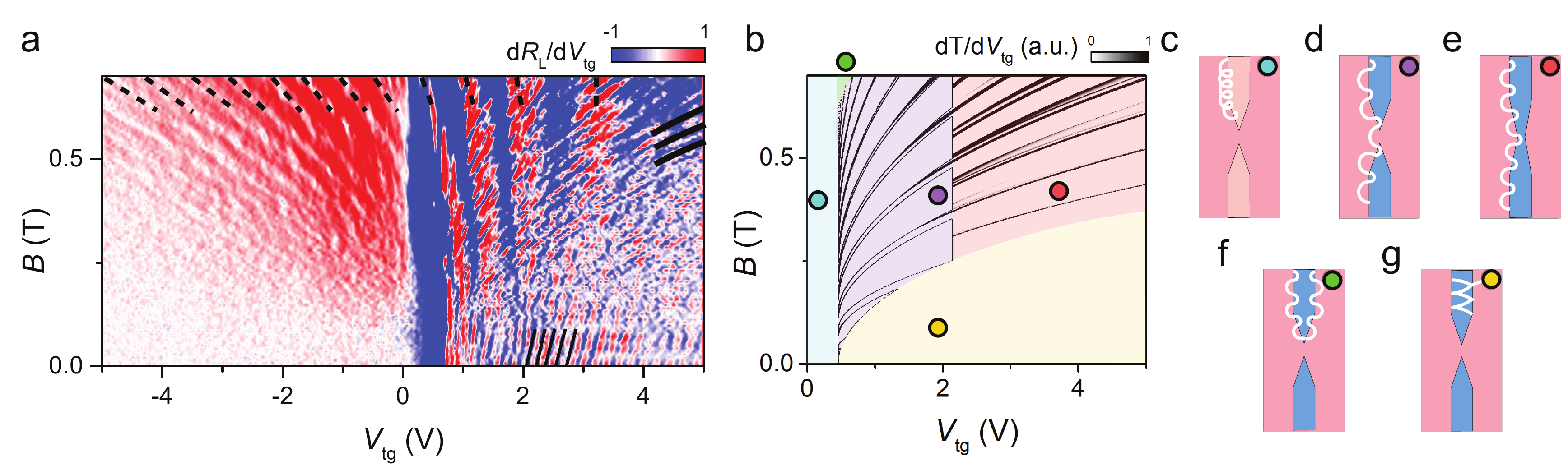}
    \caption{(a) Derivative of the longitudinal resistance d$R_{\text{L}}$/d$V_{\text{tg}}$ versus $V_{\text{tg}}$ and $B$ at $V_{\text{bg}} = -3$~V. The three sets of oscillations are indicated by guiding lines: Fabry-P\'erot oscillations (thin lines), snake states (thick lines) and Shubnikov-de-Haas oscillations from the constriction (dashed lines). (b) Numerical calculations of the snake-state transmission at the same back-gate voltage as in (a). (c-g) Schematics of the electron trajectories in the different regimes of the simulation: (c) unipolar p-p'-p regime, (d) bipolar regime with open constriction and (e) closed constriction. The regime where snake trajectories are unable to reach the other side of the constriction (f), and the Fabry-P\'erot regime discussed earlier in the text (g), give no snake-state oscillation.}
    \label{fig5}
\end{figure*}

\indent We now discuss in more details the transition with magnetic field between the ballistic conduction below the top gates and the quantum Hall edge channel conduction through the constriction. The magnetic field dependence of the FP oscillations is displayed in Fig.~\ref{fig4}a at a back-gate voltage $V_{\text{bg}}=10$~V. Upon increasing $B$, the oscillations slightly shift towards more negative $V_{\text{tg}}$ and undergo a sudden phase jump at a magnetic field of about 50~mT (and $-50$~mT). This phase jump is highlighted in Fig. 4b showing out-of-phase oscillations for line-cuts below and above the jump (see Fig.~\ref{fig5}a and Fig.~S7 of the Supplementary Information for data at other back-gate voltages). This phase jump, that has been observed in earlier works using top-gated graphene devices~\cite{Young2009, Gu2011, Nam2011, Grushina2013, Du2017, Ghahari2017}, originates from closed orbits in momentum space. At low field, cyclotron orbits within the top-gated region do not enclose the origin (see Fig.~\ref{fig4}f). Upon increasing magnetic field, trajectories bend further, until the transverse momentum changes sign, causing the orbit in k-space to enclose the Dirac point (see Fig.~\ref{fig4}e). At this particular field, electrons pick up an additional Berry phase of $\pi$, causing a sudden phase shift of the oscillation pattern. One can geometrically calculate the critical field $B_0$ at which this phase jump occurs (see Supplementary Information section 3). We find $B_0 = 57$mT using $L_{\text{eff}} = 380\,$nm, in good agreement with the experimental value.

Figure~\ref{fig4}c shows the evolution of d$R_{\text{L}}$/d$V_{\text{tg}}$ with magnetic field over a larger range of field and voltage. The graph can be divided into unipolar nn'n (right) and bipolar np'n (left) regimes separated by the Dirac point in the top-gated region at $V_{\text{tg}}=-1.5$~V. In the bipolar regime, the FP oscillations disappear at a field of about 0.25~T, and new oscillations appear at higher field.
The vanishing of FP oscillations corresponds to the field $B^* = \hbar k_{\text{F}}/e L_{\text{eff}} $ at which the cyclotron radius equals the effective length $L_{\text{eff}}$ of the top-gated cavity. At $V_{\text{bg}}=10$~V and $V_{\text{tg}}=-3$~V, we calculate $B^* \simeq 0.24$~T using $L_{\text{eff}}=380$~nm, in excellent agreement with the data. 

Above $B^*$, the cyclotron radius is smaller than the cavity length, and the electron trajectories, which are close to normal incidence due to Klein collimation, do not reach the opposite side of the cavity. Interference no longer occurs, resulting in the fading of the oscillations, similarly to what was already reported in the case of a single p-n junction \cite{Rickhaus2015}. Simultaneously, snake orbits, formed by half cyclotron orbits of opposite chirality in the n and p regions, appear along the p-n interface\cite{Taychatanapat2015,Rickhaus2015,Makk2018}. In our split-gate geometry, the snake orbits follow the boundary of the upper top-gate, hence guiding the electrons through the constriction (Fig.~\ref{fig4}d). In the unipolar regime, snake trajectories also exist at the n-n' (or p-p') interface, with a different shape, and can also drive electron trajectories through the constriction (see Supplementary Information section 7). The field $B^*$ therefore marks the onset of electron transport through the constriction.

Above 0.25~T, a second set of oscillations emerges (Fig.~\ref{fig4}c), which disperse with magnetic field and bend toward higher positive top-gate voltages. 
These oscillations correspond to those observed in Fig.~\ref{fig3}c,d, that is, to the Shubnikov-de-Haas oscillations of the Landau levels inside the constriction (see Supplementary Information section 5). Since the charge neutrality in the constriction is obtained at $V_{\rm tg}=-8$~V (for this back-gate voltage), the $N=0$ Landau level is outside the graph on the left side, and the graph diplays only the right part of the fan diagram. Interestingly, this fan does not show the usual straight lines spreading out from the charge neutrality point, but have instead a finite spacing at zero magnetic field and a finite curvature. This unusual Landau level spectrum is the direct consequence of the confinement of the cyclotron orbits in the saddle potential as explained previously (see also Supplementary Information section 6).

The guiding of the electron trajectories through the constriction at finite magnetic field relies on the existence of snake states at the top-gate pn interface. Signatures of these snake trajectories are visible in Fig. \ref{fig4}c (red arrows at the top left) as a set of weak oscillations crossing the strong SdH oscillations discussed above, and dispersing in the opposite direction. In the range of voltage and field where these oscillations are visible, the constriction is open, but the cyclotron diameter of the snake orbits is larger than the width of the constriction, such that the trajectories can either pass through or jump over the constriction, resulting in conductance oscillations.

To investigate the shape and properties of these oscillations, we focus on another set of data taken at a smaller back-gate voltage that enables us to reach the charge neutrality point in the constriction (at the maximum negative top-gate voltage accessible) and explore density profiles changing from an open constriction to a continuous barrier. The resistance map at $V_{\rm bg}=-3$~V is presented in Fig.~\ref{fig5}a (note the reversed voltage polarity, corresponding to hole carriers in the bulk regions).
We observe the same features as in Fig.~\ref{fig4}c, namely Fabry-P\'erot oscillations at low field, Landau levels from the constriction at high field, and a weak set of snake oscillations marked by thick black lines, which are very similar to those reported in single pn junction~\cite{Williams2011,Taychatanapat2015,Rickhaus2015,Makk2018}. At $V_{\rm tg}>2$~V, the constriction is indeed closed and the density profile is similar to the case of a continuous top gate. In this situation, the snake trajectories traveling along the first pn interface can either end on the injector side, leading zero transmission, or enter into the inner region of the junction and eventually escape on the collector side after traveling along the second pn interface, leading non-zero transmission.

To confirm the snake state origin of the oscillations in our split-gated device, we performed numerical calculations in which the end point of the snake trajectory is calculated\cite{Taychatanapat2015}, in order to determine if the snake trajectory ends on the injector side, or on the collector side, for instance passing through the constriction (see Supplementary Informations section 7).

The result of the calculations (for the same parameters as in Fig.~\ref{fig5}a) are presented in Fig.~\ref{fig5}b. In the bipolar p-n-p regime ($V_{\text{tg}}>0.46$~V), the snake oscillations have a similar shape as in single p-n junctions, with a $r_c^{tg} \propto \sqrt{n_{tg}} = \sqrt{C_{\text{tg}} \, V_{\text{tg}} + C_{\text{bg}}\, V_{\text{bg}}}$ dispersion, as they are described by the same condition $r_c^{bg} + r_c^{tg} \propto \sqrt{n_b} + \sqrt{n_{tg}} =$~constant (with $r_c^{bg}$ and $r_c^{tg}$ the cyclotron radius in the bulk graphene and top-gated region, respectively) \cite{Taychatanapat2015}.
Several regimes can be distinguished, that correspond to different relative sizes of $r_c^{bg}$ and of the constriction width, and to different polarities in the constriction with respect to the bulk and top-gated regions (see Fig.~\ref{fig5}c-g).
In the unipolar regime (see Fig.~\ref{fig5}c), snake oscillations are extremely weak. This could be the consequence of a strong reduction of the oscillation's amplitude due to their much longer trajectories which are very sensitive to disorder effects\cite{Taychatanapat2015}. 
Increasing $V_{\text{tg}}$, one reaches the bipolar regime with an open constriction (see Fig.~\ref{fig5}d). In this regime, snake trajectories are either transmitted through the constriction or reach the other top-gated area and leads to transmission at the bottom edge of the graphene flake. Increasing further $V_{\text{tg}}$, the constriction closes (see Fig.\ref{fig5}e) and the system becomes equivalent to a single pnp junction. At high magnetic field, the cyclotron orbit becomes smaller than the constriction size, so that all orbits go through the constriction and the oscillations should disappear (see Fig.~\ref{fig5}f). In the low field regime (yellow area in Fig.~\ref{fig5}b), electrons are transmitted through the top-gated regions (with FP oscillations) and no snake state appears (see Fig.~\ref{fig5}g).

The calculation qualitatively reproduces the snake oscillations observed in the bipolar regime in Fig.~\ref{fig5}a.
Between $V_{\text{tg}} = 0.46$~V and $V_{\text{tg}} = 2.14$~V, the constriction is open, and we experimentally observe oscillations in this regime, which further confirms our interpretation in terms of snake states driving electron trajectories through the constriction. This is also the case in the data at $V_{\text{bg}} = 10$~V (Fig.~\ref{fig4}c), where small-amplitude oscillations can be observed when the constriction is open (in the whole top-gate range). The calculation shows less oscillations than what is experimentally observed, which is probably due to the simplicity of our model, which in particular does not account for the roughness of the top-gate edges, that could change the incidence angle on the second p-n junction and increase the number of oscillations observed. We also note that disorder should decreases the visibility of the snake oscillations \cite{Taychatanapat2015}. Since the mean free path of about 1~$\mu$m is shorter than the gate length, disorder effects probably explain the small visibility of our snake oscillations.

In conclusion, we reported a thorough study of the evolution with magnetic field of an electrostatically-defined constriction in high-mobility graphene. At low magnetic field, Fabry-P\'{e}rot oscillations within the ballistic top-gated region develop and exhibit at a particular magnetic field a phase shift due to the graphene Berry phase. At higher magnetic field, the Shubnikov-de-Haas oscillations reveal that transport takes place through the constriction. The non-constant Landau level spacing is consistent with the confinement effect induced by the saddle potential in the constriction. We identify the magnetic field $B^*$ at which the transition between the two regimes occurs, that is, when the cyclotron radius becomes smaller than the length of the Fabry-P\'{e}rot cavity. 
This interpretation is supported by the observation of snake states features at intermediate magnetic field, confirmed by numerical simulations. Our findings show that signatures of the split-gate defined saddle potential can emerge at relatively low magnetic field due to the guiding of the electrons towards the constriction via snake trajectories.
Our study provides a complete understanding of the functioning of a split-gate-defined constriction in monolayer graphene from zero magnetic field up to the quantum Hall regime.

\textbf{Acknowledgments} This research was supported by the H2020 ERC grant \textit{QUEST} No. 637815.\\

\bibliographystyle{apsrev4-1} 
\bibliography{Veyrat_MainText}

\clearpage
\onecolumngrid
\setcounter{figure}{0}
\setcounter{section}{0}
\renewcommand{\thefigure}{S\arabic{figure}}

\begin{center}
\textbf{\large - Supplementary Information -}
\end{center}
\vspace{0mm}

\tableofcontents

\newpage
\section{Self-consistent electrostatic simulation of the split-gated graphene device}

\begin{figure}
\begin{center}
\includegraphics[width=12cm]{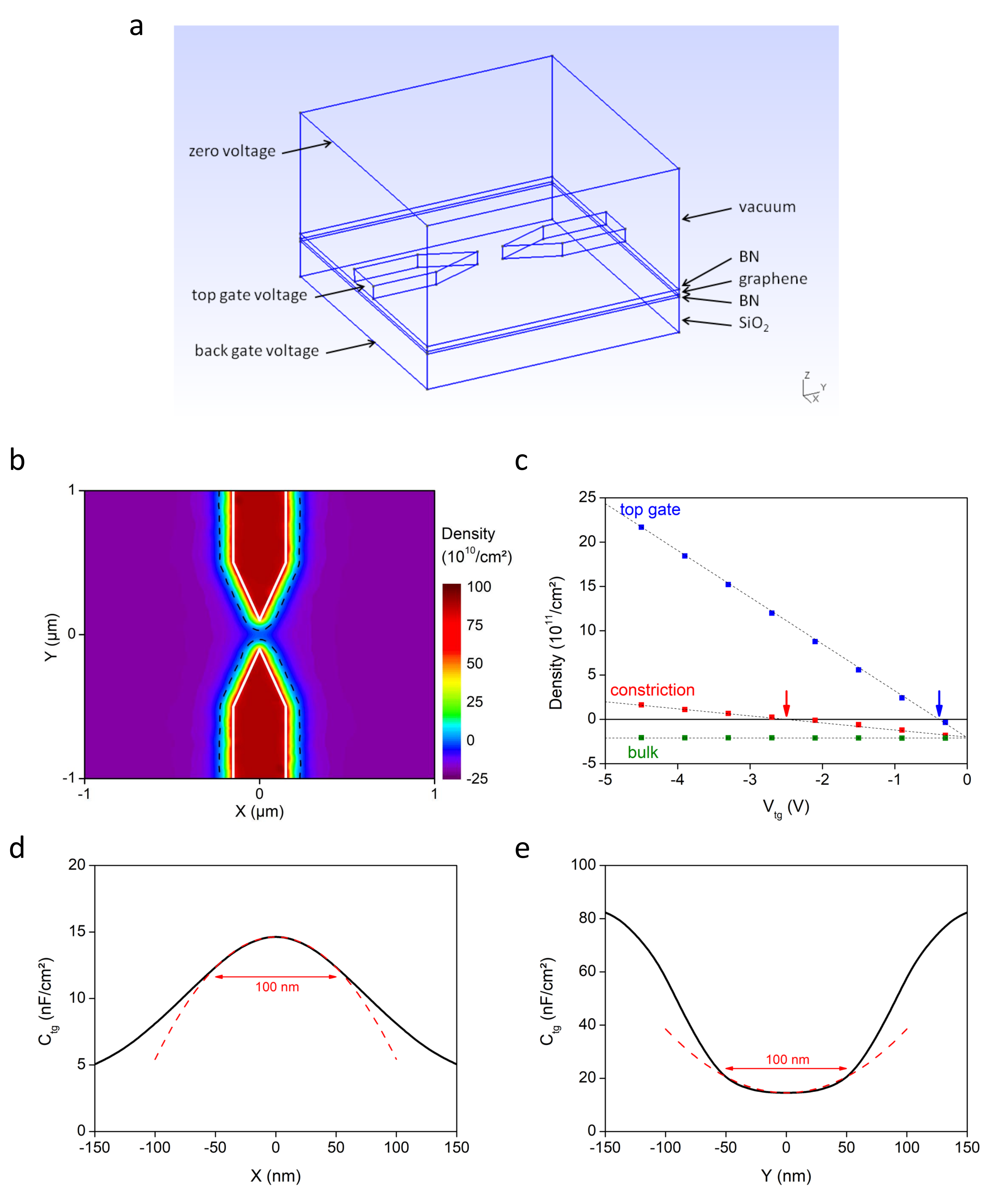}
\caption{(a) Simulation domain of size $2\times2\times1.3$ $\mu$m$^3$. The split-gate is 300~nm wide, 100~nm thick, and the gap is 200~nm wide. The SiO$_2$ layer is 280~nm thick. The bottom (top) BN layer is 20~nm (40~nm) thick. The SiO$_2$ and BN dielectric constants are set to $\epsilon_{\rm r,\,SiO_2}=3.9$ and $\epsilon_{\rm r,\,BN}=4.0$. The top surface in vacuum is set at zero voltage. The back-gate voltage is applied on the bottom surface. The top-gate voltage is applied on all faces of the split-gate volume. The zero-surface-charge boundary condition is applied on the lateral surfaces of the simulation domain. (b) Self-consistent carrier density in the graphene sheet for the split-gated geometry, at $V_{\rm bg}=3$~V and $V_{\rm tg}=-2.1$~V. The position of the zero-density curve (dashed line) indicates that the constriction contains the same type of carriers as the bulk at this gate voltages (electrons corresponds to negative density). The position of the split-gate is represented by the white line. (c) Density in the bulk of the graphene sheet (green dots), below the top gates far from the constriction (blue dots), and at the saddle-point of the constriction (red dots), for $V_{\rm bg}=3$~V and $V_{\rm tg}$ varying from 0 to $-5$~V. The blue arrow at $V_{\rm tg}=-0.4$~V indicates the charge neutrality condition below the top gates and the red arrow at $V_{\rm tg}=-2.5$~V indicates the charge neutrality condition in the constriction. (d,e) Capacitance profiles along the longitudinal $x$ direction (d) and along the transverse $y$ direction (e) extracted from the self-consistent electrostatic simulations. These profiles are fitted with parabolas between $-50$~nm and $+50$~nm to extract the curvature of the density profile in the two perpendicular directions (red dashed lines).}
\label{figureS1}
\end{center}
\end{figure}

\begin{figure}[h!]
\begin{center}
\includegraphics[width=14cm,trim={0 0 0 0},clip]{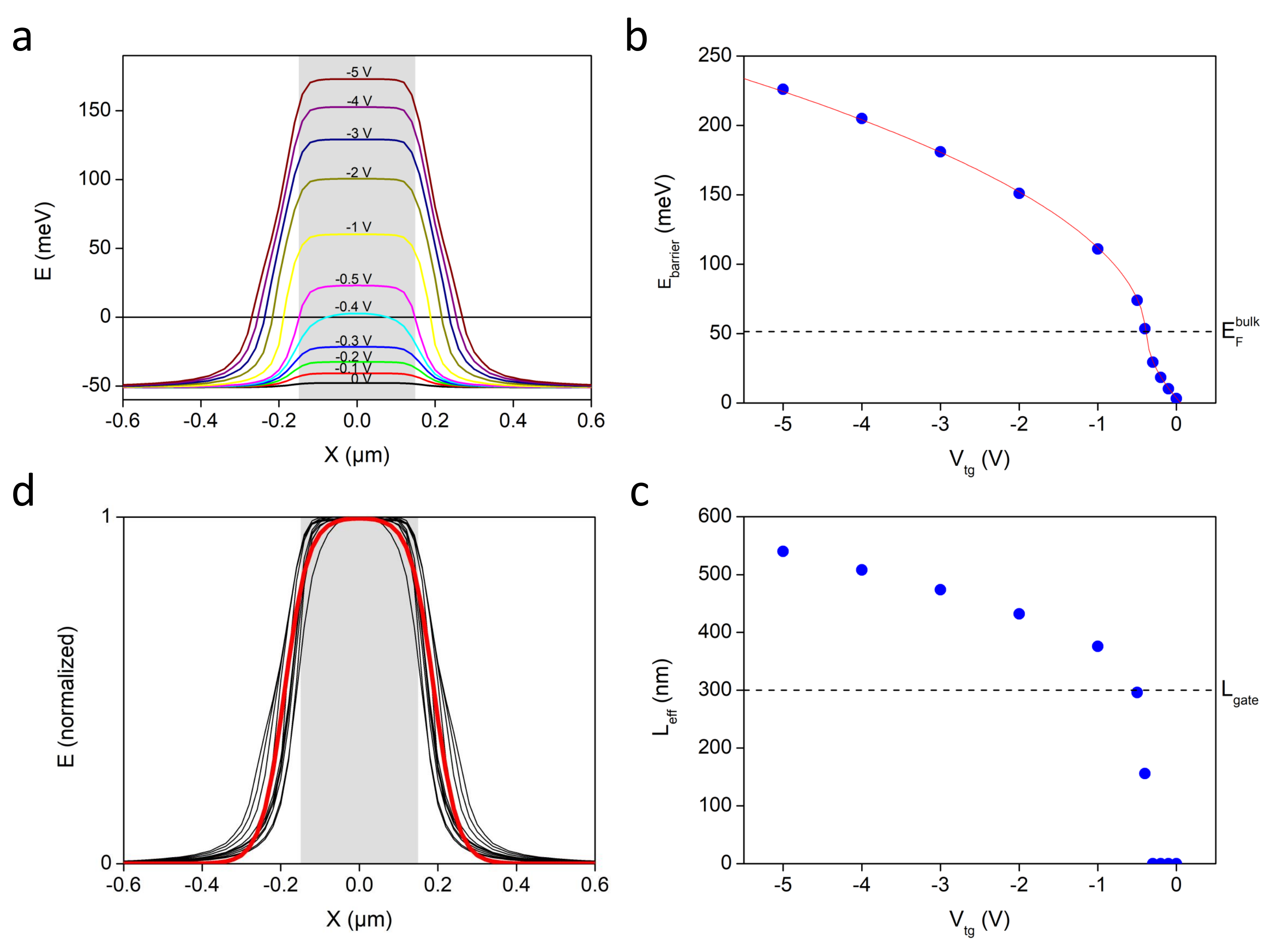}
\caption{(a) Self-consistent electrostatic energy profile $E=-eV$ across the npn junction, for a continuous top gate, at $V_{\rm bg}=3$~V and $V_{\rm tg}$ varying from 0 to $-5$~V. (b) Energy barrier between the bulk and the top-gated regions. The red line is the theoretical value calculated with the analytical expression for the quantum capacitance in graphene. (c) Cavity length defined as the distance between the two zeros of density in (b). The cavity length is larger than the gate size due to the close proximity of the top gate as compared to the back gate. (d) Normalized energy profiles (black curves) identical to those shown in (b) and analytical function (red curve) modeling the ensemble of energy profiles.} 
\label{figureS2}
\end{center}
\end{figure}

To complement and support our analysis of the split-gated graphene device, we carried out self-consistent electrostatic simulations of the carrier density in the graphene plane as a function of the back-gate and top-gate voltages. The geometry of the simulation is represented in Fig.~\ref{figureS1}a. The dimensions of the split-gate and the thicknesses of the SiO$_2$ and BN layers are close to those of the measured device. The graphene sheet is modeled by a charge density $\sigma$ linked to the electrostatic potential $V$ by the relation :
$$ \sigma = (-e)\,{\rm sign}(V)\,\frac{e^2 V^2}{\pi\hbar^2v_F^2} $$
The mesh grid is computed using Gmsh (http://gmsh.info) and the electrostatic problem is solved self-consistently using a modified version of MaxFEM (http://www.usc.es/en/proxectos/maxfem), an electromagnetic simulation software based on the finite element method.

An example of carrier density distribution within the graphene sheet is shown in Fig.~\ref{figureS1}b. The densities in the bulk, below the top gates, and at the saddle-point of the constriction, are plotted in Fig.~\ref{figureS1}c for a fixed back-gate voltage and various top-gate voltages. The capacitive couplings of the top gate are quite linear in this range of large charge densities, and the weak effects of the non-linear screening in graphene are only visible very close to zero charge density (blue and red arrows). The charge neutrality in the constriction is obtained at $V_{\rm tg}=-2.5$~V for $V_{\rm bg}=3$~V, such that the capacitance between the top gate and the constriction is 1.2 times larger than the back-gate capacitance, in correct agreement with the experimental value obtained from Fig.~3d of the article. 

The local capacitance $C_{\rm tg}(x,y)$ can be extracted from the density maps using the relation $\sigma(x,y)=-C_{\rm bg}\,(V_{\rm bg}-V(x,y))-C_{\rm tg}(x,y)\,(V_{\rm tg}-V(x,y))$ which includes the non-linear quantum capacitance of graphene \cite{Liu2013} through the relation between $\sigma(x,y)$ and $V(x,y)$ given above. The capacitance profiles around the constriction along the $x$ and $y$ directions are plotted in Fig.~\ref{figureS1}d,e and fitted with parabolas to extract the curvature coefficients $\alpha=(1/2e)({\rm d^2}C_{\rm tg,0}/{\rm d}y^2)$ and $\beta=-(1/2e)({\rm d^2}C_{\rm tg,0}/{\rm d}x^2)$ used in section~6. Over a spatial extension of 100~nm, the curvature coefficients are found to be $\alpha=1.5\times10^{29}$~m$^{-4}$V$^{-1}$ in the $y$ direction and $\beta=0.6\times10^{29}$~m$^{-4}$V$^{-1}$ in the $x$ direction.

The profile of the potential energy $E=-eV$ across the top-gated region (far from the constriction) is plotted in Fig.~\ref{figureS2}a for a fixed back-gate voltage $V_{\rm bg}=3$~V and various top-gate voltages. A subset of potential profiles at $V_{\rm tg}=-0.2$, $-0.4$, $-1$, and $-2$~V are plotted tin Fig.~1c of the article. The barrier height plotted in Fig.~\ref{figureS2}b increases non-linearly with the gate voltage as expected in graphene (red line) with a rapid variation around the Dirac point (dashed line) when switching from the nn'n to the npn configuration. Above the threshold of the bipolar regime (npn), the cavity length between the two zeros of density increases quickly to the gate size of 300~nm, and then continues to increase slowly to larger values as shown in Fig.~\ref{figureS2}c.

For the transport simulations described in the next section, the potential profile needs to be modeled by an analytical function. We use the exact potential profiles calculated at $V_{\rm bg}=3$~V and $V_{\rm tg}$ varying between 0 and $-5$~V (Fig.~\ref{figureS2}a) to obtain an averaged profile (Fig.~\ref{figureS2}d) which is then modeled by the function :
$$ f(x) = \frac{1}{4} \left(1+\tanh\left(\frac{x+d/2}{w}\right)\right) \left(1+\tanh\left(\frac{d/2-x}{w}\right)\right) $$
This modeled profile is the product of two step functions separated by a distance $d=380$~nm with a characteristic transition half-width $w=60$~nm. This value of $d$ is taken as a fixed parameter for all gate voltages during the transport simulations. Note however that the cavity length $L_{\rm eff}$ is not equal to $d$, and strongly depends on the gate voltages as discussed above.

For comparison with the experimental value of the cavity length $L_{\rm eff}=380$~nm measured at $V_{\rm bg}=10$~V and $V_{\rm tg}\approx-3$~V, we computed the potential profile at these gate voltages and obtained a cavity length of 366~nm, in good agreement with the experimental value. The electric field ${\cal E}=-{\rm d}V/{\rm d}x$ at the pn interface can also be obtained from this potential profile and we obtain ${\cal E} = 3.3\times 10^6$~V/m. This value is used in the article to calculate the collimation angle $\theta$ through the pn interface.

\newpage
\section{Simulation of the conductance through the npn junction}

\begin{figure}[b!]
\begin{center}
\includegraphics[width=16cm]{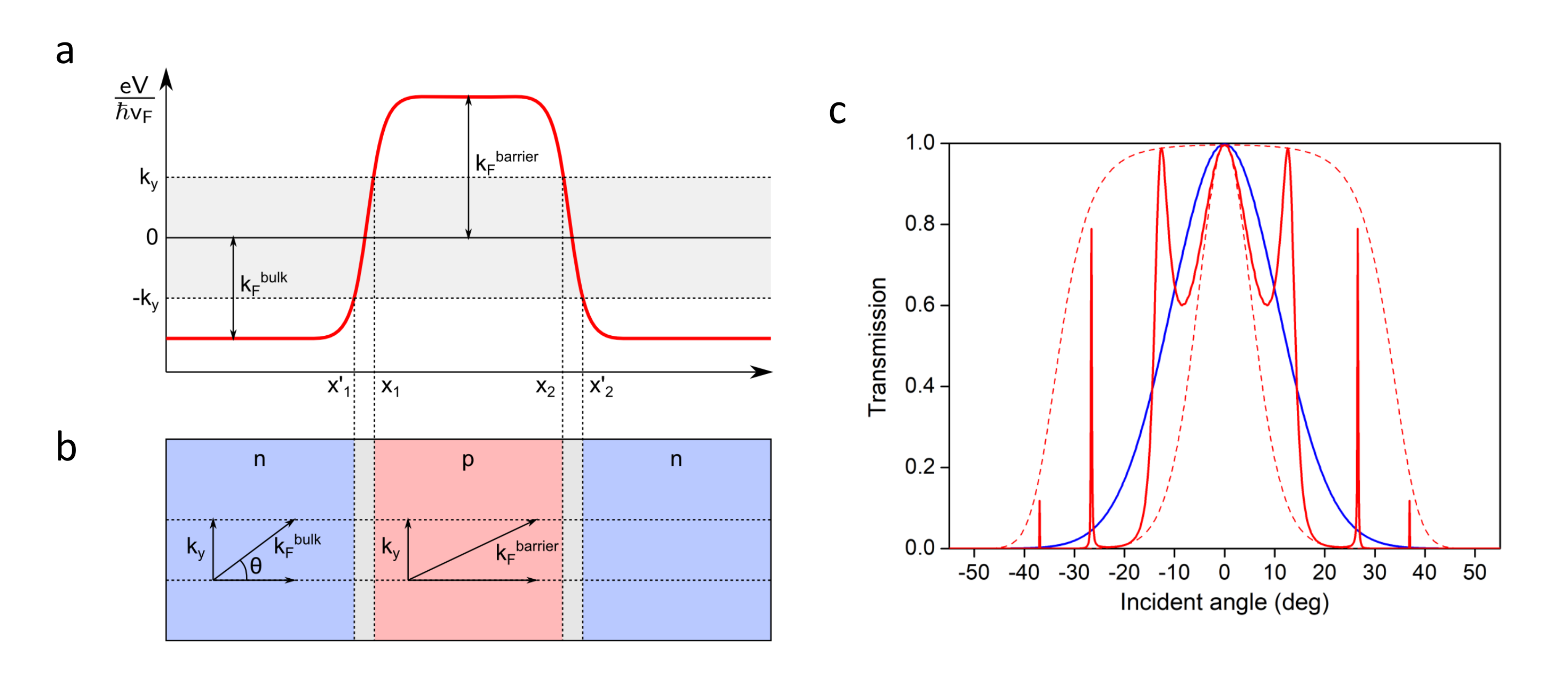}
\caption{(a) Schematics of the npn junction, with a Fermi wave vector in the n region determined by the back-gate only, and a Fermi wave vector in the p region determined by both the back-gate and the top-gate. (b) Due to conservation of the parallel wave vector $k_y$, an electron arriving at the Fermi level with incident angle $\theta$ has to cross a tunneling barrier at the pn interface (region in gray). The classical turning points $x_1'$ and $x_1$ (resp. $x_2$ and $x_2'$) for the first (second) pn interfaces are obtained as the intersections of the potential profile $k_{\rm F}(x)$ with the constant $\pm k_y$ horizontal lines. (c) Angular dependence of the transmission $T(\theta)$ calculated for $V_{\rm bg}=3$ V, $V_{\rm tg}=-1$ V, and the potential profile described in the text (red line). The red dashed lines indicated the envelope of the interference peaks ($\phi\equiv0$ and $\pi$). The blue line is the transmission $|t_1|^2$ of a single pn interface showing the Klein collimation effect.}
\label{figureS3}
\end{center}
\end{figure}

Numerical simulations of the transmission through the npn junction have been performed to compare with experimental data. For a given incident angle $\theta$ on the left pn interface, the transmission $T$ through the npn junction can be calculated in the WKB approximation \cite{Gu2011} at zero magnetic field with the formula :
$$ T(\theta) = \left|\frac{t_1t_2}{1-r_1r_2\exp(i2\phi)}\right|^2 $$
The coefficient $t_1=e^{-\lambda}$ with $\lambda=\int_{x_1'}^{x_1}(k_y^2-k_{\rm F}^2)^{1/2}{\rm d}x$ is the transmission amplitude across the pn interface which forms a tunneling barrier between the turning points $x_1'$ and $x_1$ (Fig.~\ref{figureS3}a). The coefficient $r_1=(1-e^{-2\lambda})^{1/2}$ is the reflection amplitude at the pn interface from the inside of the npn cavity. The quantity $\phi=\int_{x_1}^{x_2}(k_{\rm F}^2-k_y^2)^{1/2}{\rm d}x$ is the phase accumulated in the central region between the turning points $x_1$ and $x_2$. At zero magnetic field, the two pn interfaces have the same transmission and reflection coefficients $t_1=t_2$ and $r_1=r_2$.

The parallel wave vector $k_y(\theta)=k_{\rm F}^{\rm bulk}\sin(\theta)$ is conserved and fixed by the incidence angle $\theta$ (Fig.~\ref{figureS3}b), whereas the total Fermi wave vector $k_{\rm F}(x)=k_{\rm F}^{\rm bulk}+(k_{\rm F}^{\rm barrier}-k_{\rm F}^{\rm bulk})\times f(x)$ is a function of the position $x$ across the npn junction. The Fermi wave vectors $k_{\rm F}^{\rm bulk}$ and $k_{\rm F}^{\rm barrier}$ are obtained from the density $n$ within each region by $k_{\rm F}=\sqrt{\pi n}$. The function $f(x)$, that characterizes both the energy profile $E_{\rm F}(x)$ and the wave vector profile $k_{\rm F}(x)=E_{\rm F}(x)/\hbar v_{\rm F}$, is modeled by the product of two step functions separated by a fixed distance $d=380$~nm with a characteristic transition half-width $w=60$~nm (see details in the previous section).

The angular dependence of the transmission $T(\theta)$ shows a series of resonant peaks corresponding to constructive interference (red curve in Fig.~\ref{figureS3}c). The presence of a tunneling barrier at the pn interface is responsible for the Klein collimation effect with a small range of incident angles being transmitted (blue curve). The resonant peak around $\theta=13\deg$ corresponds to half transmission of the pn interface and gives the main contribution to the conductance oscillations. 

The conductance $G$ of the npn junction is obtained by averaging the angular dependence of the transmission according to the formula :
$$ G = \frac{4e^2}{h} \; \frac{Wk_{\rm F}^{\rm bulk}}{\pi} \int_{-\pi/2}^{\pi/2} T(\theta) \; \frac{{\rm d}\theta}{\pi} $$
where $W$ is the width of the sample. The choice of a uniform averaging over the incident angle $\theta$ is justified by the fact that all angles are almost equiprobable for our sample geometry (see Fig.~1a in the article), as opposed to experiments on nanoribbons with transverse quantization of the incident wave vector.

The conductance oscillations calculated as a function of the top-gate and back-gate voltages are shown in Fig.~2d of the article. They reproduce correctly the experimental pattern of resistance oscillations in Fig.~2c. Interestingly, the curvature of the interference fringes at low back-gate voltage corresponds to a significant increase of the cavity length, which results from a weaker screening of the top-gate voltage by the low carrier density in the bulk region.

\newpage
\section{Magnetic field $B_0$ of the Berry phase shift}

In the main text, we show that Fabry-Perot oscillations undergo a phase shift at a field $B_0$. The field at which Farby-Perot oscillations undergo a phase shift due to the geometrical Berry phase can be calculated as $B_0\simeq\hbar k_{\text{F}}\theta/eL_{\text{eff}}$, where $\hbar$ and $e$ are the reduced Planck constant and electron charge respectively, $k_{\text{F}}$ is the Fermi wave-vector in the central region and $\theta$ is the typical incidence angle of the trajectories contributing the most effectively to the FP interference \cite{Young2009, Grushina2013}. The typical incidence angle on the barrier emerges from the strong angular dependence of the tunneling transmission through a smooth pn junction (Klein collimation) \cite{Cheianov2006PRB}. It can be calculated as the angle for which the transmission is $1/2$, giving $\theta=\sqrt{\ln(2)e\mathcal{E}/\pi\hbar v_{\rm F}k_{\rm F}^2}$ where $\mathcal{E}$ is the electric field at the pn junction. Self-consistent simulations for $V_{\text{bg}}=10$~V and $V_{\text{tg}}=-3$~V ($n=p=6.2\times10^{11}$~cm$^{-2}$) give the value $\mathcal{E}=3.3\times 10^6$~V/m (see Section 1 of this Supplementary Information), which results in $\theta=13^\circ$ and $B_0=0.057$~T using $L_{\rm eff}=380$~nm. The good agreement with the experimental value of $B_0$ confirms the Berry phase origin of the observed phase shift. This expression of $B_0$ is valid for an angular transmission dominated by small angles, which is the case in our device (see Section 2 of this Supplementary Information). The small value of $\theta$ justifies the linearized formula for $B_0$.


\newpage
\section{Landau level fan diagram in the homogeneous graphene regions}

\begin{figure*}[h!]
    \centering
    \includegraphics[width=0.8\linewidth]{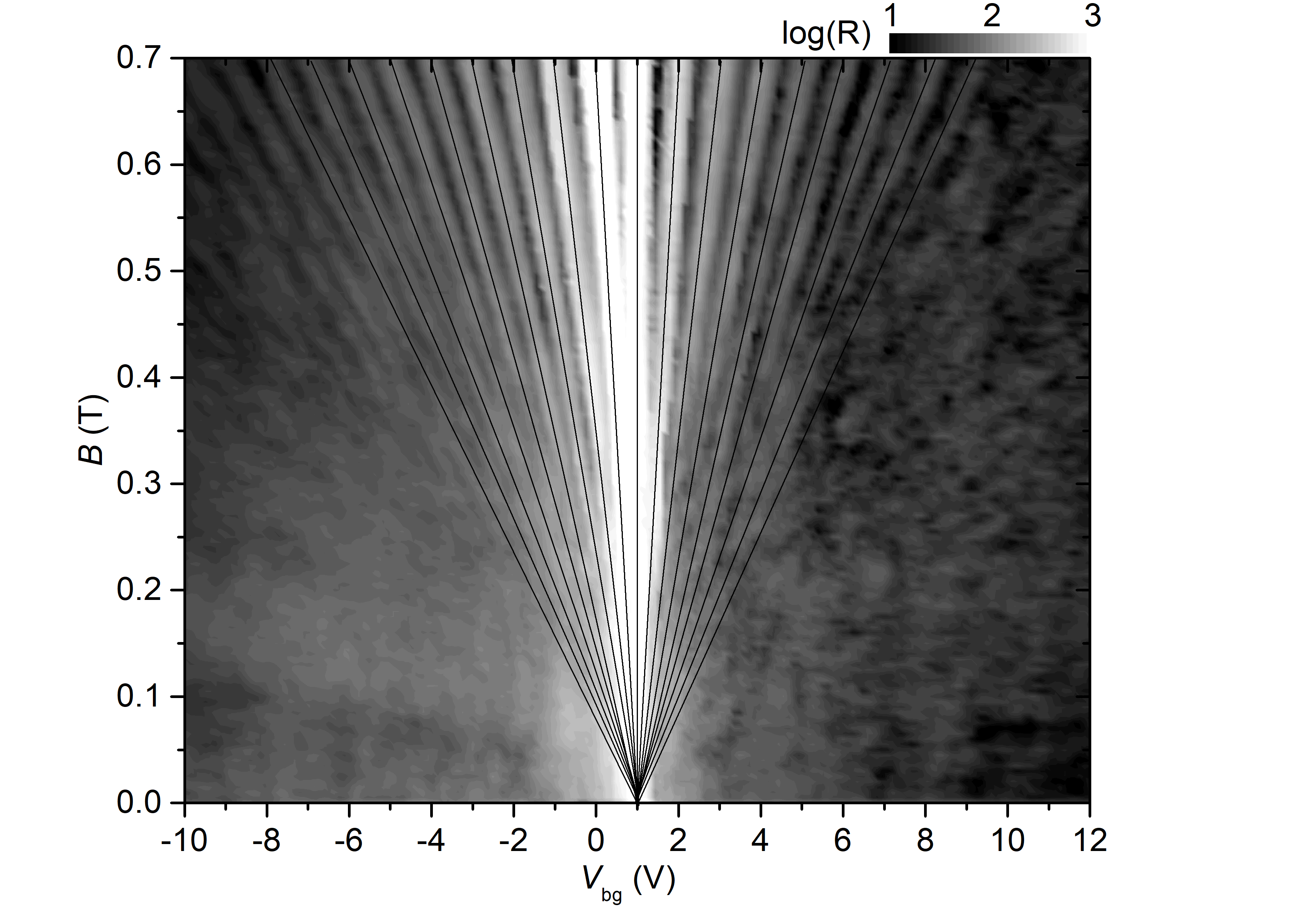}
    \caption{Magnetoresistance versus back-gate voltage at low magnetic field (at T = 4.2 K), measured in the homogeneous part of the device. The black lines are guide for the eye showing the linear dispersion of the bulk Landau levels.}
    \label{fig_Fan}
\end{figure*}

Although deviations from a linear Fan diagram are observed for the Landau levels from the constriction, a standard Fan diagram is expected for the homogeneous graphene. In Fig.~\ref{fig_Fan}, we present a Fan diagram where the four-probe resistance is measured with the two voltage probes located on the same side of the device (to avoid a possibly non-uniform density below the top gates). The Fan diagram is linear, as expected in the absence of a confining/deconfining potential. 

\newpage
\section{Correspondence between Landau levels in ($V_{\text{tg}}$, $V_{\text{bg}}$) and ($V_{\text{tg}}$, $B$) maps}

\begin{figure*}[h!]
    \centering
    \includegraphics[width=0.5\linewidth]{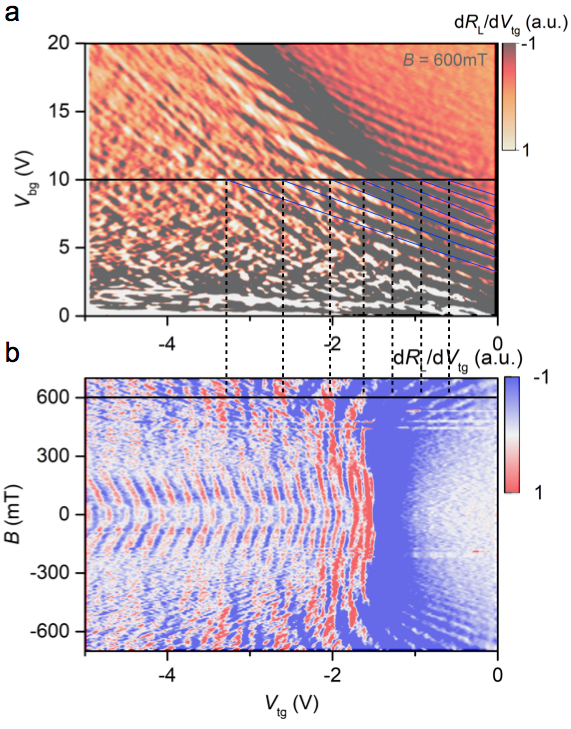}
    \caption{(a) Derivative of the longitudinal resistance d$R_{\text{L}}$/d$V_{\text{tg}}$ versus $V_{\text{bg}}$ and $V_{\text{tg}}$ at $B=600$~mT. (b) Derivative of the longitudinal resistance d$R_{\text{L}}$/d$V_{\text{tg}}$ as a function of the magnetic field $B$ and $V_{\text{tg}}$, at $V_{\text{bg}}=10$~V. In both panels, the black line corresponds to the same conditions $V_{\text{bg}}=10$~V and $B=600$~mT. The Shubnikov-de-Haas oscillations from the constricted area are located by the dotted lines. The blue lines are guide for the eye locating the minima of the oscillations.}
    \label{fig_oscillations}
\end{figure*}

Figure~\ref{fig_oscillations} shows the comparison between the ($V_{\text{tg}}$, $V_{\text{bg}}$) map of d$R_{\text{L}}$/d$V_{\text{tg}}$ at $B=600$~mT, and the ($V_{\text{tg}}$, $B$) map at $V_{\text{bg}}=10$~V (respectively from Fig.~3 and Fig.~4 of the main text). The black line in both figures correspond to the same gates and field condition. The dashed black lines identify the SdH oscillations from the constriction in both panels. In Fig.~\ref{fig_oscillations}b, these oscillations correspond the right part of a fan diagram (they all disperse to more positive $V_{\text{tg}}$ with increasing $B$), as expected since the carrier polarity in the constriction does not change over the top-gate voltage range, at this back-gate voltage. The deviations from a linear fan diagram result from the confinement in the saddle potential of the constriction.

\newpage
\section{Landau level spectrum in a saddle-point potential}

In Fig.~3d of the article (recorded at $B=1$~T), the spacing between the Landau levels is constant versus back-gate voltage, but this spacing changes when varying the top-gate voltage. This dependence on top-gate voltage is evidenced in Fig.~\ref{figureLL}a (same data as in Fig.~3d) where the red lines, passing through each Landau level at $V_{\rm tg}=0$ and parallel to the $N=0$ Landau level, do not follow the experimental Landau level position (black/white lines for negative/positive top-gate voltage). This deviation from a constant Landau level spacing results from the non-uniform potential landscape in the constriction, where the combined actions of the back gate and the top gate create a saddle potential. In the following, we present theoretical models that take into account this particular potential landscape in the calculation of the Landau level spectrum. The good agreement between the observed Landau level spacing and our theoretical models provides another evidence that the current flows through the constriction above $B^*$.

The theoretical expression for Landau levels in a saddle potential has been calculated for semiconductor 2DEGs with a quadratic dispersion relation \cite{Fertig87,Buttiker90}, but not for graphene, since there is no analytical solution in case of a linear dispersion relation, except in the high magnetic field limit \cite{Floser2010}. In the following, we first consider a potential which is non-uniform in the transverse ($y$) direction only, and then a saddle potential with opposite curvatures in the transverse ($y$) and longitudinal ($x$) directions.

\subsection{Landau levels in a one-dimensional parabolic potential}

The first situation is modeled by a region with a non-uniform electron density $n(y)=n_0-ay^2$ in the transverse direction, placed in a uniform perpendicular magnetic field $B$. By choosing a gauge such that $\vec{A}=-By\vec{u}_x$, the system is invariant along the $x$-axis and the momentum $p_x$ is conserved. For the non-uniform density considered here, trajectories with non-zero $p_x$ are drifting along the $x$-axis, whereas trajectories with $p_x=0$ form closed cyclotron orbits. Following the approach given in Ref.~\cite{Gu2011}, the semi-classical Bohr-Sommerfeld quantization condition $\oint p_y\,dy=N2\pi\hbar$, for a closed orbit with quantum number $N$ (no $1/2$ zero-point energy term in graphene), then writes :
\begin{equation}
\int_{y_1}^{y_2} \sqrt{(\hbar k_F)^2 - (eBy)^2}\,dy = N\pi\hbar
\end{equation}
where $y_1$ and $y_2$ are the classical turning points that cancel the term in the integral, and $k_F(y)=\sqrt{\pi n(y)}$ is the non-uniform Fermi wave vector. For the parabolic density profile considered here, this equation gives equidistant levels as a function of the density $n_0$ :
\begin{equation}
n_{0,N} = N \sqrt{(4eB/h)^2 + 4a/\pi}
\label{LL1}
\end{equation}
In this expression, the term $4eB/h=gB/\phi_0$ is the electron density per Landau level in a uniform graphene sheet, with the factor $g=4$ corresponding to the four-fold spin and valley degeneracies in graphene. The second term $4a/\pi$ gives the modification of the level spacing due to the parabolic density profile governed by the curvature parameter $a$. This expression holds both for a confinement potential with $a>0$ and for a barrier potential with $a<0$, up to the critical situation $a=-(\pi/4)(4eB/h)^2$ called ``Landau level collapse'' where the level spacing drops to zero \cite{Gu2011}.

\begin{figure}[b!]
\begin{center}
\includegraphics[width=17cm,trim={0 0 0 0},clip]{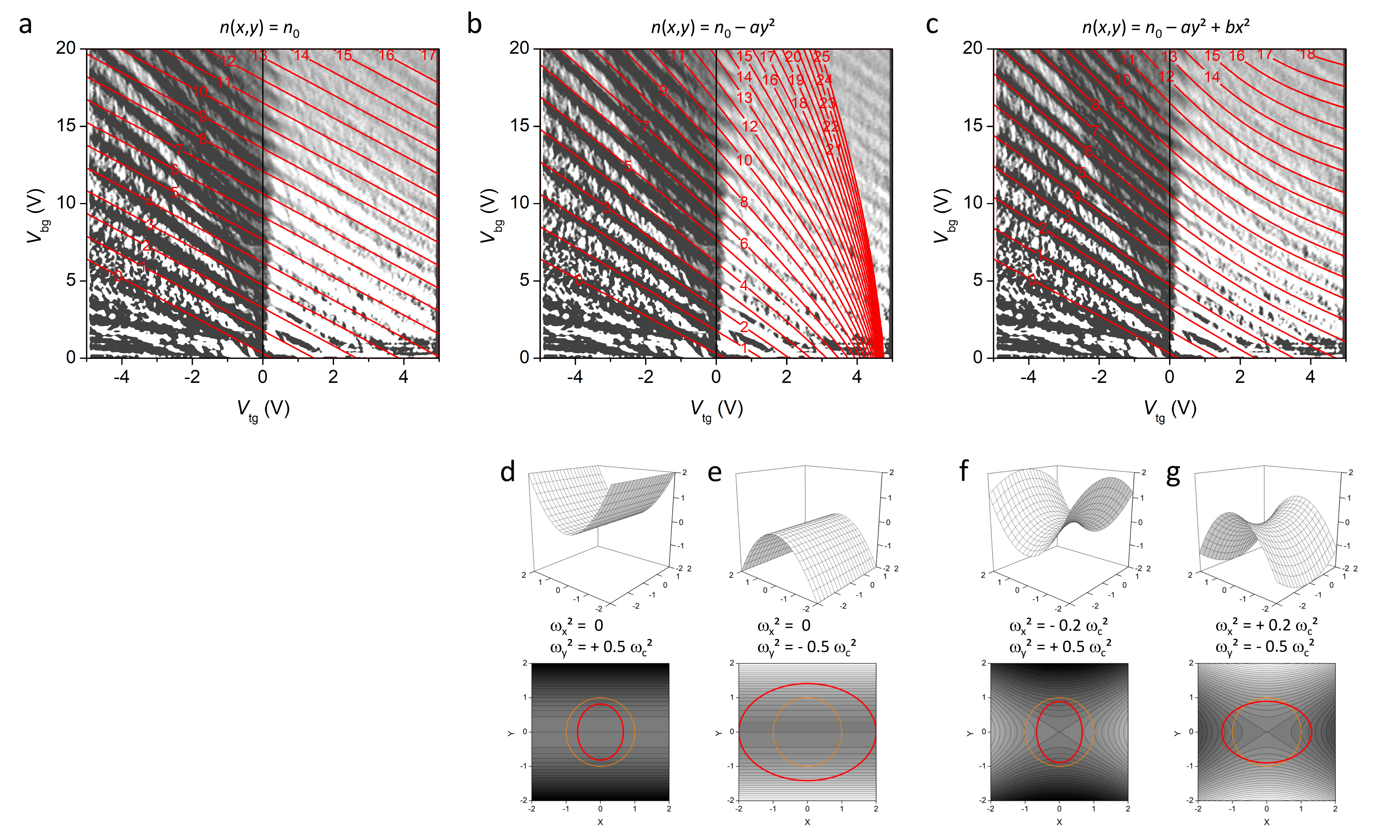}
\caption{(a-c) Same data as in Fig.~3d at 1~T (greyscale plot) compared to the expected Landau level spectrum in the constriction (red lines) for a uniform density (a), a parabolic density (b), and a saddle density (c). The density contains a uniform part controlled by the back-gate voltage $V_{\rm bg}$ and a non-uniform part controlled by the split-gate voltage $V_{\rm tg}$. The Landau level spacing is constant in (a), changes with top-gate voltage in (b) and (c), and shows a collapse in (b) for large deconfinement potentials. The red numbers indicate the Landau level index $N$. The red lines in (c) are the same as those plotted in Fig.~3d. The validity of the theoretical expression used in (c) is limited to split-gate voltages below 4~V. (d-g) Cyclotron orbits (red lines) in parabolic potentials (d,e) and saddle potentials (f,g) for semiconductor 2DEGs with Hamiltonian $H=\frac{1}{2m}(p_x-eBy/2)^2+\frac{1}{2m}(p_y+eBx/2)^2+\frac{1}{2}m\omega_x^2x^2+\frac{1}{2}m\omega_y^2y^2$. The curvature coefficients $\omega_x^2$ and $\omega_y^2$ are expressed in terms of the cyclotron angular frequency $\omega_c=eB/m$, and they can be positive or negative, for confinement or deconfinement potentials, respectively. The spatial coordinates are in units of the cyclotron radius $r_c=\sqrt{2mE}/eB$, which is the radius of the circular orbit obtained in the uniform case (orange circles).} 
\label{figureLL}
\end{center}
\end{figure}

We now compare the position of the Landau levels given by this expression with the experimental data (Fig.~\ref{figureLL}b). Neglecting the weak non-linear quantum capacitance effects for the large densities considered here, the local electron density is simply proportional to the gate voltages and given by the local capacitance. At the center of the constriction, the density is then given by $n_0=(1/e)(C_{\rm bg}V_{\rm bg}+C_{\rm tg,0}V_{\rm tg})$ and the curvature parameter is given by $a=-(1/2e)({\rm d^2}C_{\rm tg,0}/{\rm d}y^2)V_{\rm tg}=-\alpha V_{\rm tg}$ (the minus sign is only valid for positive $V_{\rm bg}$). In the left part of the map where $V_{\rm tg}<0$, and above the line with zero-density in the constriction, the curvature is positive ($a>0$) and corresponds to an enhanced confinement of the cyclotron orbits (larger Landau level spacing). In the right part of the map where $V_{\rm tg}>0$, the curvature is negative ($a<0$) and corresponds to a deconfinement of the cyclotron orbits (smaller Landau level spacing). On the vertical line of the map at $V_{\rm tg}=0$, the density is uniform ($a=0$) and the Landau level spacing is unperturbed with $n_{0,N}=N(4eB/h)$. We use this situation to finely tune the value $C_{\rm bg}$ of the back-gate capacitance such as to reproduce the observed Landau level spacing. Assuming electron-hole symmetry, the $N=0$ Landau level should also remain unperturbed, and we use this situation to finely tune the value $C_{\rm tg,0}$ of the top-gate capacitance in the middle of the constriction.

By fitting the Landau level pattern with this expression in the left part of the map (Fig.~\ref{figureLL}b, for $V_{\rm tg}<0$), one gets a curvature coefficient $\alpha=1.5\times10^{29}$~m$^{-4}$V$^{-1}$, in good agreement with the self-consistent electrostatic simulations of the split-gated device, giving the same value for the transverse density profile fitted over a spatial range of 100~nm around the saddle point (see Section 1 and Fig.~\ref{figureS1}e). The simulated density profile is not strictly parabolic over the large spatial extension of a cyclotron orbit whose diameter $d_c=2l_B\sqrt{2N}$ (for graphene) is several times the magnetic length $l_B$ (26~nm at 1~T). This cyclotron diameter is about 100~nm for the level index $N=2$ and reaches 200~nm for $N=8$ in the top part of the map. The influence of this non-parabolicity of the density profile is beyond the model presented here and would require a detailed theoretical study.

\subsection{Landau levels in a saddle potential}

In the right part of the map (Fig.~\ref{figureLL}b, for $V_{\rm tg}>0$), this model with a non-uniform density in only one direction predicts a collapse of the Landau level spacing which is not observed in the experiment. The reason is the presence of a saddle potential in the constriction, with a weak confinement potential in the longitudinal direction, which restores a finite level spacing when this one would have been collapsed by the large deconfinement potential in the transverse direction for positive $V_{\rm tg}$ (and positive $V_{\rm bg}$). Physically, the Landau level collapse corresponds to an opening of the cyclotron trajectory, and even a weak confinement potential in the perpendicular direction is enough to restore a closed trajectory. This effect is explained in Fig.~\ref{figureLL}d-g showing the classical cyclotron orbits obtained for parabolic potentials and saddle potentials in semiconductor 2DEGs where the theoretical equations can be solved analytically. Confinement potentials (d) reduce the orbit length and therefore increase the Landau level spacing (b, left). Deconfinement potentials (e) increase the orbit length and therefore decrease the Landau level spacing (b, right). When the closed cyclotron orbits breaks into open trajectories, the Landau level spectrum collapses (b, extreme right). For saddle potentials with large confinement (f), the small perpendicular deconfinement plays a little role, because the orbit is narrow in the deconfining direction (lines in b and c are similar on the left side). In contrast, for saddle potentials with large deconfinement (g), the small perpendicular confinement plays an important role, because the orbit is wide in the confining direction (lines in b and c are very different on the right side). In this last situation, the orbit length (g) is strongly reduced as compared to (e), and is similar to the orbit length of the uniform case (orange circle), restoring the initial Landau level spacing.

For a parabolic saddle potential, the Landau level quantization cannot be solved analytically in graphene where the dispersion relation is linear. For this reason, we use the exact result obtained for a quadratic dispersion relation \cite{Fertig87,Buttiker90}, and we modify it to obtain an approximate expression for graphene. As pointed out in Ref.~\cite{Gu2011}, the semi-classical quantization condition takes the same form for particles with linear and quadratic dispersion relations having the same spatial distribution of density (but obviously not the same distribution of potential). The only differences are the degeneracy term $g=4$ or 2, and the zero-point energy term $\gamma=0$ or 1/2, for mass-less and massive carriers, respectively. Using this analogy and the relation $n(x,y)=(gm^*/2\pi\hbar^2)\times(E_F-U(x,y))$ for semiconductor 2DEGs, we convert the existing analytical expression for the energy spectrum $E_{0,N}$ into a density spectrum $n_{0,N}$ (the 0 index refers to the saddle point values), and then put the graphene values of $g$ and $\gamma$ to obtain the approximate density spectrum in a graphene constriction with density $n(x,y)=n_0-ay^2+bx^2$ : 
\begin{equation}
n_{0,N} = N \sqrt{\frac{ \sqrt{ \Big((4eB/h)^2+4a/\pi-4b/\pi\Big)^2 + 64ab/\pi^2 } + \Big((4eB/h)^2+4a/\pi-4b/\pi\Big) }{2}}
\label{LL2}
\end{equation}
For $b=0$, this expression recovers the exact semi-classical result obtained above for the graphene linear dispersion relation, showing the relevance of using the analogy between graphene and semiconductor 2DEGs with the same density profile.

In the classical capacitance model introduced before, the curvature parameters in the transverse and longitudinal directions can be written as $a=-\alpha V_{\rm tg}$ and $b=-\beta V_{\rm tg}$, respectively, with positive $\alpha$ and $\beta$ coefficients (for $V_{\rm bg}>0$). By fitting the Landau level pattern in both parts of the map with this expression (Fig.~\ref{figureLL}c), one gets curvature coefficients $\alpha=1.5\times10^{29}$~m$^{-4}$V$^{-1}$ and $\beta=0.6\times10^{29}$~m$^{-4}$V$^{-1}$. This $\beta$ value corresponds exactly to the longitudinal curvature of the density profile obtained by self-consistent electrostatic simulations (see Section 1 and Fig.~\ref{figureS1}d), and the $\alpha$ value corresponds to the transverse curvature fitted over a spatial range of 100~nm (the density is not strictly parabolic in this direction) corresponding to the cyclotron diameter at $N=2$. This good agreement gives confidence in the validity of our interpretation in terms of simultaneous confinement and deconfinement effects around a saddle point. The above expression is however an approximation for graphene, obtained by analogy with semiconductor 2DEGs, and is \textit{a priori} not the exact expression. Interestingly, the same quality of fit and the same coefficients $\alpha$ and $\beta$ are obtained when fitting the Landau level pattern with the second-order Taylor expansion of the above expression with respect to the curvatures $a$ and $b$ (with a linear and a quadratic terms), meaning that this expression and an hypothetical exact solution for graphene would not differ before the third order terms.

Finally, we note that the exact quantum solution found in Ref.~\cite{Fertig87} for semiconductor 2DEGs assumes that their quantity $\Omega$ is positive, corresponding here to $(4eB/h)+2(4a/\pi-4b/\pi)>0$. In the case of a large deconfinement potential, the negative term can dominate over the positive sum of the cyclotron and confinement terms, and the exact solution cannot be applied anymore. Using the above values of $\alpha$ and $\beta$, this limit occurs at $V_{\rm tg}=4$~V for $B=1$~T, and it has been taken into account in Fig.~3d by interrupting the lines marking the calculated Landau levels position. This limit has also been taken into account in Fig.~\ref{fig_fitLLB} below.

\subsection{Landau level fan diagram in a saddle potential}

\begin{figure*}[b!]
    \centering
    \includegraphics[width=0.6\linewidth]{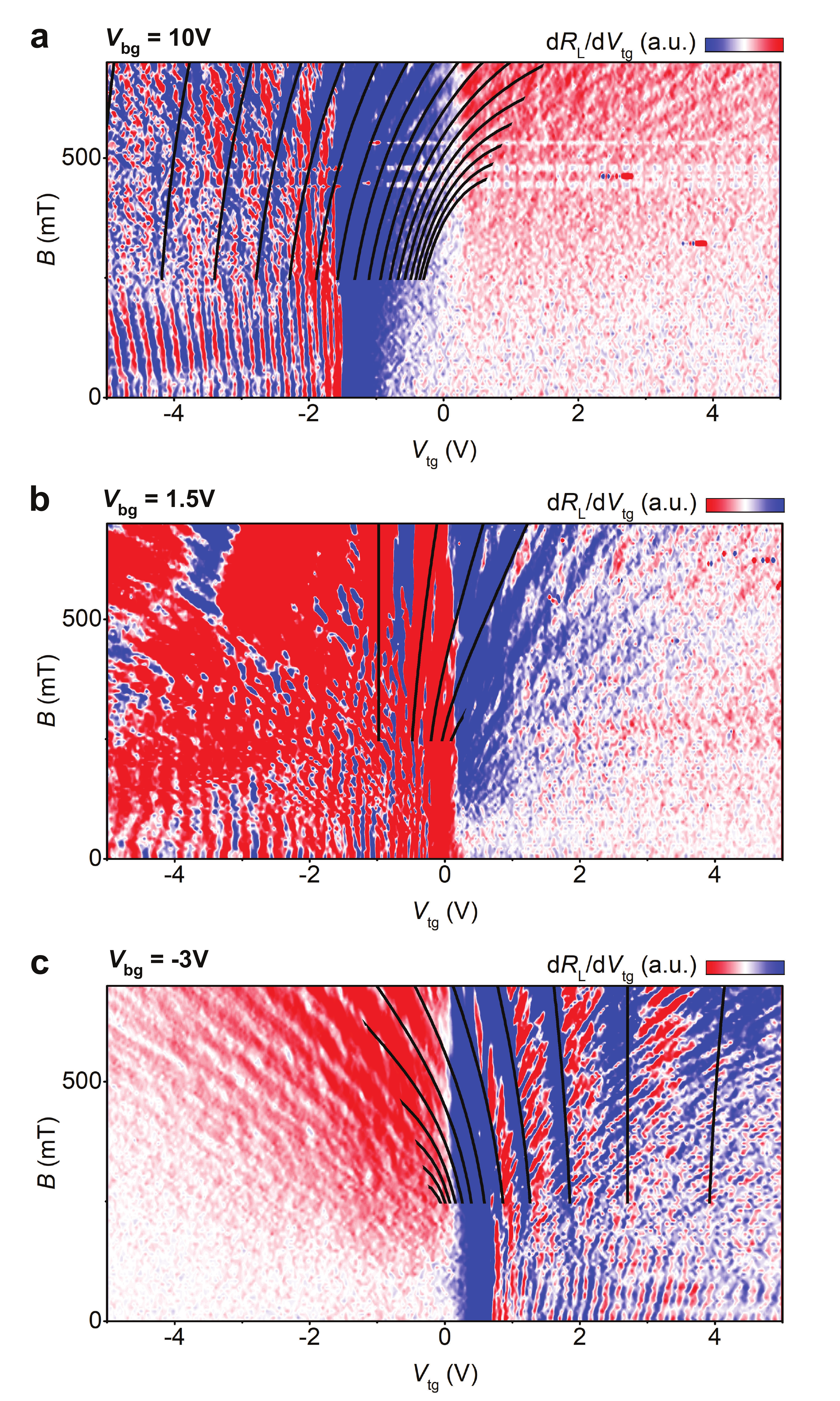}
    \caption{Derivative of the longitudinal resistance d$R_{\text{L}}$/d$V_{\text{tg}}$ versus $V_{\text{tg}}$ and $B$ at (a) $V_{\text{bg}} = 10$~V, (b) $V_{\text{bg}} = 1.5$~V, and (c) $V_{\text{bg}} = -3$~V. The calculated Landau levels in the constriction are shown by black lines. The Landau levels are only represented above the critical field $B^*$ (250~mT) which marks the onset of the conduction through the constriction. In the unipolar regime, at high top-gate voltage, the approximate expression (Eq.~(3)) is not valid (see text) and has therefore not been plotted.}
    \label{fig_fitLLB}
\end{figure*}

In a uniform potential, the evolution of the Landau level spectrum versus magnetic field and carrier density draw a linear fan diagram, both for graphene and semiconductor 2DEGs. Here, the fan diagram in the constriction is not linear, but instead shows a set of curved lines due to the confinement potential in the constriction. In Fig.~\ref{fig_fitLLB}, we show the comparison between the experimental derivative of the resistance, and the fitted Landau levels in the constriction, marked by black lines, at different back-gate voltages. We find an excellent agreement, which shows the robustness of our interpretation. The best fit leads to the same value for $\beta$ as found above at $B=1$~T (Fig.~3d and Fig.~\ref{figureLL}c), and a slightly larger value $\alpha=1.8\times10^{29}$~m$^{-4}$V$^{-1}$, corresponding to a smaller curvature. This reflects the fact that the potential curvature must be considered over a larger range than $100$~nm (as done at 1~T), owing to the larger cyclotron orbits at the lower fields considered here (see Fig.~\ref{figureS1}e).

\newpage
\section{Snake trajectories in a split-gated device}

\subsection{Snake trajectories in the unipolar regime}

\begin{figure}[b!]
\begin{center}
\includegraphics[width=0.8\linewidth]{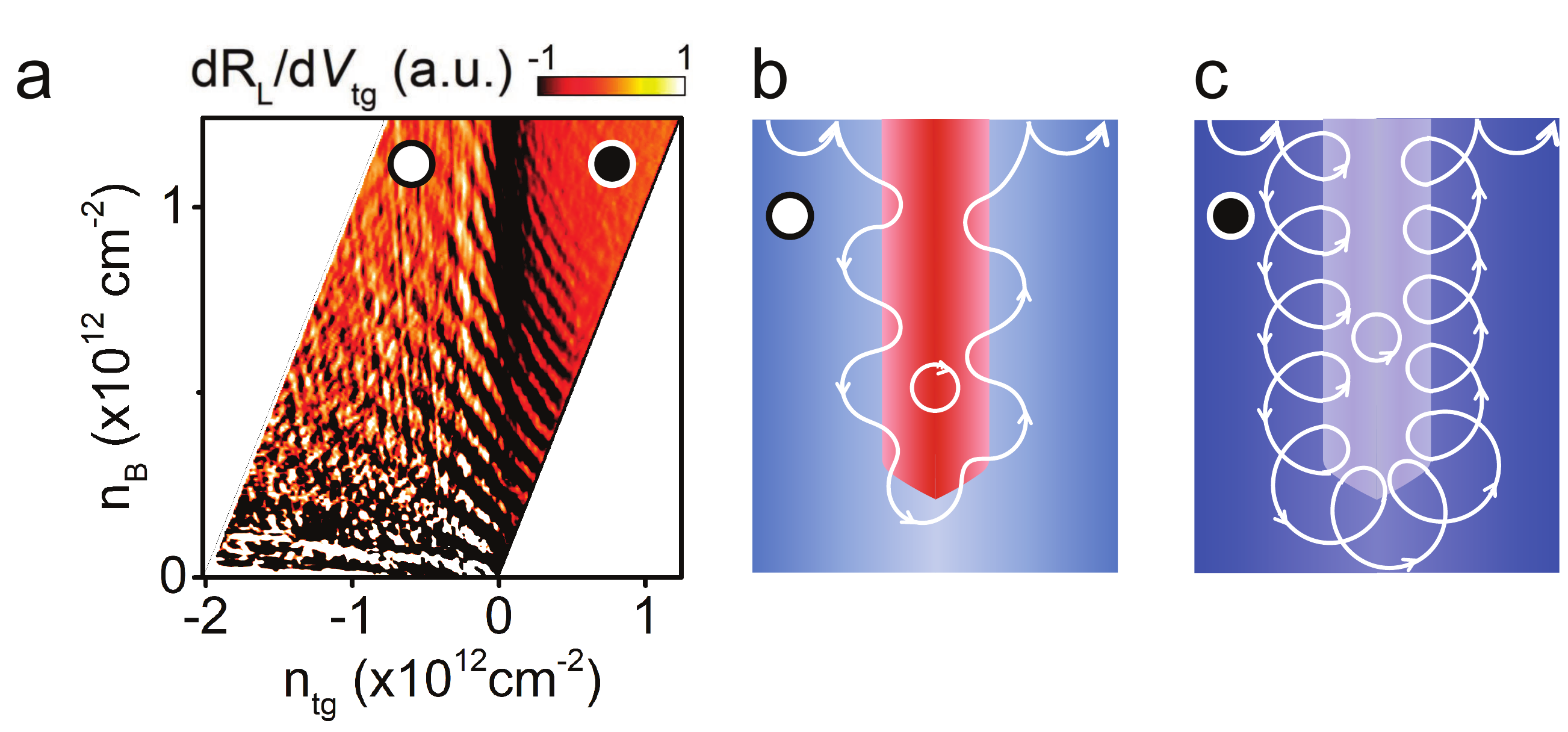}
\caption{(a) Derivative of the longitudinal resistance d$R_{\text{L}}$/d$V_{\text{tg}}$ versus $n_{\text{tg}}$ and $n_{\text{B}}$ (the carrier densities of the top-gated region and of the bulk graphene, respectively) at $B=600$~mT. The white and black dots refers to the schematics (b) and (c) of the snake states in the bipolar and unipolar regimes of the upper top-gated region.}
\label{figureS8}
\end{center}
\end{figure}

When increasing the magnetic field above $B^*$, the FP oscillations disappear and are replaced by resistance oscillations due to snake orbits. In this regime, the cyclotron orbits drift along the edges of the split-gate electrodes, and the electrons are guided towards the constriction.

In the bipolar regime, the snake trajectories are made of half-circular orbits with opposite chirality in the n- and p-doped regions (Fig.~\ref{figureS8}b), while in the unipolar regime, the two half-circular orbits have the same chirality but different cyclotron radii (Fig.~\ref{figureS8}c). Both shapes of trajectories drive the electrons through the constriction, so that the feature of the Landau levels of the constricted region are still visible in the unipolar regime.

The snake trajectories of the unipolar regime appear for a cyclotron radius $r_c < L_{\text{eff}}$. This happens for densities in the top-gated region $n_{\text{tg}}$ such that $n_{\text{tg}} < \frac{1}{\pi} \left( \frac{L_{\text{eff}} \, eB}{\hbar} \right)^2 = 3.8\times 10^{12} \text{cm}^{-2}$ at $B=600$~mT. This is the case over the whole gate range presented in Fig.~3c of the main text, and plotted in Fig.~\ref{figureS8}a versus the densities in the bulk and top-gated regions.

We observed no signature of snake states in the unipolar regime. As stressed in the main text, we attribute this to disorder effects. The length of the snake trajectories is much longer in the unipolar case (Fig.~\ref{figureS8}b-c), so that these trajectories are more sensitive to disorder effects, which are known to reduce the visibility of the oscillations \cite{Taychatanapat2015}.

\subsection{Simulation of the transmission of snake states}

\begin{figure*}[t!]
    \centering
    \includegraphics[width=0.7\linewidth]{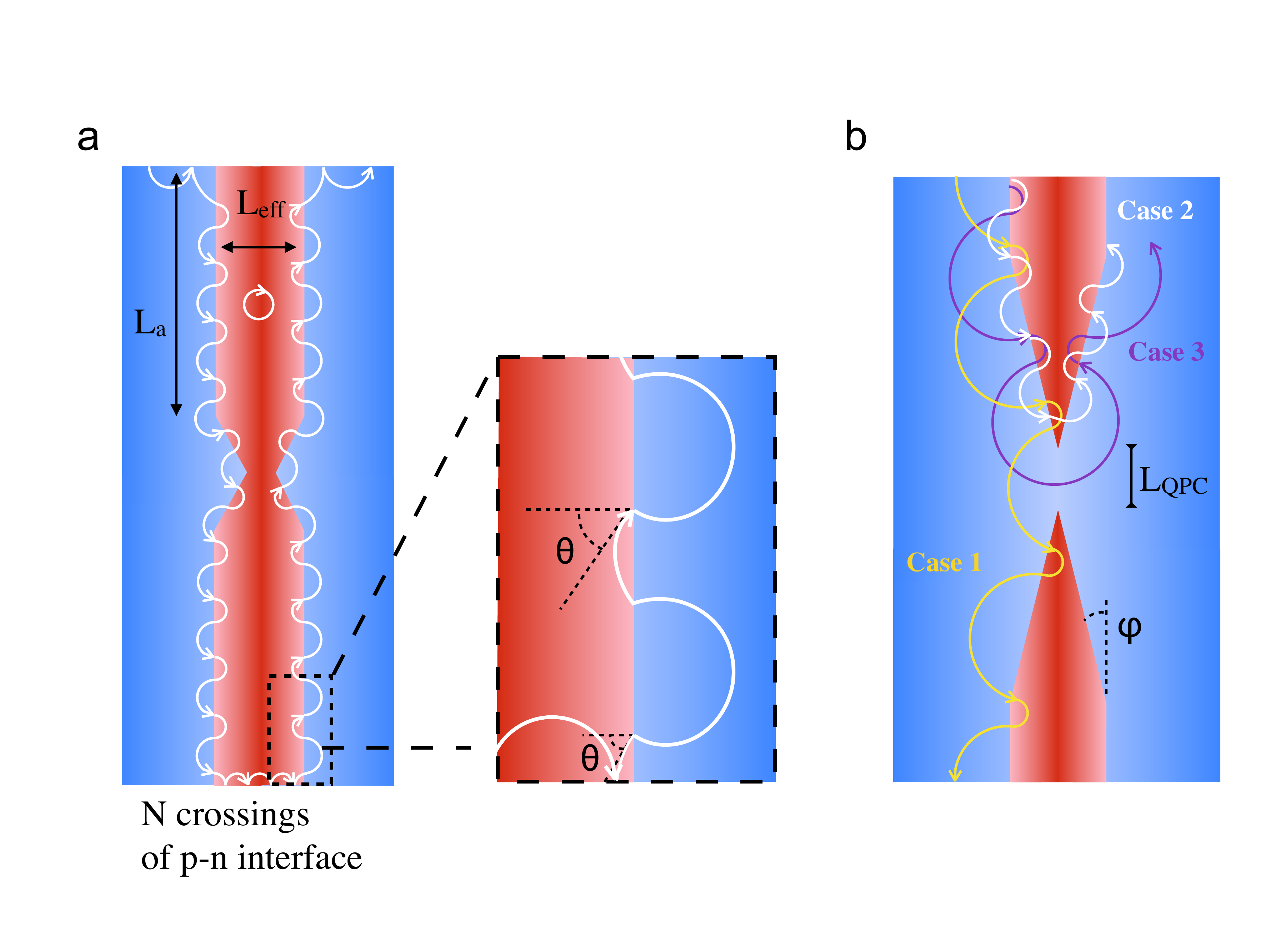}
    \caption{Schematic of the geometrical model adopted for the simulation of snake states. (a) Case of a closed constriction. In our device and simulations, $L_a=1.6$~$\mu$m, $L_{eff}=380$~nm, and $\varphi=15^o$. The enlarged figure illustrates the incidence angle $\theta$ of the snake trajectory on the second p-n interface. (b) Case of an open constriction. In our simulation, $L_{QPC}=200$~nm. The three depicted trajectories correspond to those discussed in the text.}
    \label{snake_model}
\end{figure*}

To simulate qualitatively the transmission in presence of snake trajectories, we build a model based on geometrical considerations, as done in \cite{Taychatanapat2015}. We consider a single starting point for the snake state, and calculate the final position of the trajectory for a given set of gate voltages. We discuss here only the bipolar regime, the unipolar regime being discussed in the previous section. We consider a snake state starting in the upper left corner, and crossing the first p-n interface at normal incidence, as depicted in Fig.~\ref{snake_model}a. 

For a closed constriction, the distance to reach the lower edge of the device is $L = 2L_a + 2L_{eff}\,\cos(\varphi)$, with $L_a$ the length of the straight part of the p-n interface, $L_{eff}$ the effective width of the gated area, and $\varphi$ the angle of the gate apex (see Fig.~\ref{snake_model}b). For $L_0 = N(2r_c^{bg} + 2r_c^{tg}) + r_c^{bg} $, with $N$ is the integer part of $L/(2r_c^{bg} + 2r_c^{tg})$, the end of the snake is on the injector side if $L<L_0$, leading to $T=0$, or on the top-gate side if $L>L_0$. In this last case, skipping orbits follow the graphene edge below the top gate, until the snake reaches the second p-n interface. Depending on the incidence angle $\theta$ on the second p-n interface (see Fig.~\ref{snake_model}b), the transmission of the snake is taken as $t(\theta)^N$, with $N$ the number of transmission through the second p-n interface and $t(\theta)$ the angular transmission through the p-n junction. To simplify the calculation, we use $t=\cos^2(\theta)$. $N$ being large (typically 10 to 20), only nearly normal incidence through the second p-n interface will result in non-zero transmission of the snake state.

For an open constriction, we assume that the constriction length $L_{QPC}$ varies linearly with $V_{\text{tg}}$ between 200~nm at the charge neutrality point below the top-gates (open constriction) and 0~nm at the charge neutrality point in the constriction (closed constriction), which reproduces qualitatively the effect of the change in the constriction size. We distinguish three different situations for the end point of the snake in the upper top-gated region, before reaching the constriction. We note $L' = L_a + L_{eff}\,\cos(\varphi)$, $L_1 = N(2r_c^{bg} + 2r_c^{tg}) + r_c^{bg} + 2r_c^{tg} $ where $N$ is the integer part of $L'/(2r_c^{bg} + 2r_c^{tg})$, and $l = L' - L_1$. If the two conditions (1) $L_1 < L'$ and (2) $2r_c^{bg} > L_{QCP} + 2l\,\cos(\varphi)$ are fulfilled, the snake exits the top-gate when approaching the constriction, reaches the opposite side of the constriction, and continues along the p-n interface of the lower top-gate (case 1 in Fig.~\ref{snake_model}b). In this case, the transmission is calculated as in the closed constriction case. Otherwise, if condition (1) is not fulfilled, the snake enters below the top-gate when approaching the constriction and is transmitted through the constriction with $T=1$ (case 2 in Fig.~\ref{snake_model}b). Finally, if condition (1) is fulfilled, but not condition (2), the snake cannot reach the opposite side of the constriction (case 3 in Fig.~\ref{snake_model}b) and is transmitted through the constriction with $T=1$.

\subsection{Snake-states oscillations in the biplar regime}

\begin{figure*}[t!]
    \centering
    \includegraphics[width=1\linewidth]{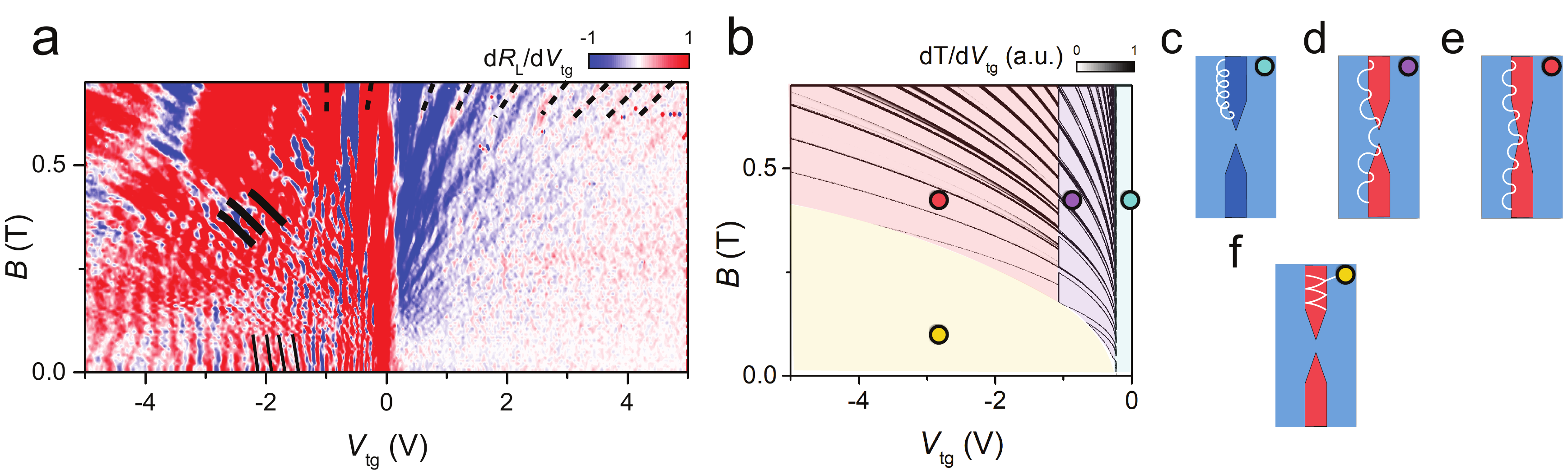}
    \caption{(a) Derivative of the longitudinal resistance d$R_{\text{L}}$/d$V_{\text{tg}}$ versus $V_{\text{tg}}$ and $B$ at $V_{\text{bg}} = 1.5$~V. The three sets of oscillations are indicated by guiding lines: Fabry-P\'erot oscillations (thin lines), snake states (thick lines) and Shubnikov-de-Haas oscillations from the constricted region (dashed lines). (b) Numerical calculations of the transmission of the snake states versus $V_{\text{tg}}$ and $B$ at $V_{\text{bg}} = 1.5$~V. (c-f) Schematics of the junction in the different zones of the simulation: unipolar n-n'-n regime (c), bipolar regime with open constriction (d) and closed constriction (e). (f) Fabry-P\'erot regime discussed in the main text.}
    \label{figS6}
\end{figure*}

Additionally to the analysis presented in the main text at $V_{\rm bg}=-3$~V (Fig.~5), we present in Fig.~\ref{figS6}a and b, the dependence of d$R_{\text{L}}$/d$V_{\text{tg}}$ versus $V_{\text{tg}}$ and $B$, and the corresponding simulations of snake oscillations, at $V_{\text{bg}}=1.5$~V. We observe the same features as in Fig.~2c, namely Fabry-P\'erot oscillations at low field, and Landau levels from the constriction at high field. Additionally, another set of oscillations appear in the bipolar regime, above $B^*$. Those oscillations are very similar to the snake oscillations reported in single p-n junctions\cite{Williams2011,Taychatanapat2015,Rickhaus2015,Makk2018}. The result of the calculations at $V_{\text{bg}} = 1.5$~V are presented in Fig.~\ref{figS6}b (see details about the modelisation above). The calculations show oscillations with a similar shape as in single p-n junctions, and qualitatively reproduce the additional set of oscillations observed at $V_{\text{bg}}=1.5$~V. The schematics in Fig.~\ref{figS6}c-f represents the different regimes of the simulations, similarly to those of Fig.~5 from the main text.

The vertical lines between the open and closed constriction regimes (at $V_{\text{tg}} \sim -1$~V) are an effect of the constriction closure. At lower gate voltage, the corresponding snake trajectories goes through the constriction, with transmission 1. At higher gate voltage, the constriction is closed, and the transmission is calculated as for a continuous n-p-n junction. This does not take into account the fact that, although closed, the constriction is at first narrow, so that the snake trajectory could nevertheless go straight to the other side as in the open constriction case with $r_c^{bg} < L_{QPC}$. A tight binding model would be necessary to remove this apparent discontinuity. Still, the visibility of the snake oscillations is experimentally too low to resolve the crossing between the open and closed constriction cases, so that our model reproduces the general shape of the experimental data.

A more detailed modelisation could be obtained by including the potential barrier, which would give the exact angular transmission of the p-n junction. It could also be improved by adding disorder. However our toy model is enough to show that the oscillations occurring in the bipolar regime in Fig.~\ref{figS6}a and b are consistent with snake states oscillations.

\bibliographystyle{ieeetr}
\bibliography{supplementary.bib}

\end{document}